\documentclass[a4paper, 12pt]{article} 
\usepackage[right=1.1in,left=1.1in,top=1.1in,bottom=1.1in]{geometry}
\usepackage{graphicx} 
\usepackage{braket}
\usepackage{mathrsfs}
\usepackage[T1]{fontenc} 
\usepackage{amsmath}
\usepackage{pxfonts}
\usepackage[T1]{fontenc}
\usepackage{amssymb}
\usepackage{natbib}
\usepackage{epstopdf}
\usepackage{setspace}
\usepackage{enumitem}
\usepackage{sectsty}
\usepackage{wrapfig} 
\sectionfont{\large}
\bibpunct{(}{)}{,}{a}{}{,}
\usepackage{tcolorbox}
    \usepackage{pgfplots}

\usepackage{subcaption}
\usepackage{adjustbox}

\usepackage{tikz}   
\usepackage{tikz-3dplot} 

\usepackage[protrusion=true,expansion=true]{microtype} 

\newcommand{\qed}{\nobreak \ifvmode \relax \else
      \ifdim\lastskip<1.5em \hskip-\lastskip
      \hskip1.5em plus0em minus0.5em \fi \nobreak
      \vrule height0.75em width0.5em depth0.25em\fi}

\usepackage{bbm}
\usepackage{MnSymbol}
\usepackage[all]{xy}

\usepackage{xcolor,colortbl,array,amssymb}

\usepackage{epigraph}
\makeatletter

\newlength{\bibitemsep}
\newlength{\bibparskip}
\let\oldthebibliography\thebibliography
\renewcommand\thebibliography[1]{%
  \oldthebibliography{#1}%
  \setlength{\parskip}{\bibitemsep}%
  \setlength{\itemsep}{\bibparskip}%
}

\newcommand{\x}[1]{{#1}}

\title{\textbf{Laws of Physics}\footnote{Prepared for \textit{Cambridge Elements in Philosophy of Physics}}} 

\author{Eddy Keming Chen\thanks{Department of Philosophy,  University of California, San Diego, 9500 Gilman Dr, La Jolla, CA 92093-0119. Website: www.eddykemingchen.net. Email: eddykemingchen@ucsd.edu  }}

\date{Draft of \today}

\begin{document}
\bibliographystyle{apa}

\maketitle 

\begin{abstract}
Despite its apparent complexity, our world seems to be governed by simple laws of physics. This volume provides a philosophical introduction to such laws.  I explain how they are connected to some of the central issues in philosophy, such as ontology, possibility, explanation, induction, counterfactuals, time, determinism, and fundamentality. I suggest that laws are fundamental facts that govern the world by constraining its physical possibilities. I examine three hallmarks of laws---simplicity, exactness, and objectivity---and discuss whether and how they may be associated with laws of physics. 
\end{abstract}

\begingroup
\singlespacing
\tableofcontents
\endgroup


\section{Introduction}


Much work in physics has been devoted to the discovery of its true fundamental laws: the basic principles that govern the world.\footnote{I use ``fundamental laws,'' ``laws of physics,'' ``physical laws,'' and ``laws'' interchangeably unless noted otherwise.}  The collection of all such laws may be called the axioms of the final theory of physics or the Theory of Everything (TOE). The fundamental laws cannot be explained in terms of deeper principles \citep[p.18]{steven1992dreams}. We use them to explain observed phenomena, including the formation of galaxies, the collisions of black holes, the stability of matter, the tidal periods of ocean waves, and the melting of ice cubes. 

Laws are intimately connected to many long-standing philosophical issues, such as modality, explanation, causation, counterfactuals, time, induction, and determinism. For example, physical possibility and necessity can be defined in terms of laws; laws contribute to scientific explanations of natural phenomena; laws support counterfactuals, predictions, and retrodictions; laws are linked to the direction of time; determinism and indeterminism are properties of laws; and so on. Anyone interested in those issues would benefit from some understanding of laws. 

There are interesting puzzles about laws themselves: 
\begin{enumerate}
  \item \textit{Metaphysical issues:} What kind of things are laws? Most people believe that laws are different from material entities such as particles and fields, because, for one thing, laws seem to \textit{govern} the material entities. But what is this governing relation? What makes material entities respect such laws? 
  \item \textit{Epistemological issues:} How do we have epistemic access to laws? Many different candidate laws can yield the same data, a phenomenon known as empirical equivalence.   How should we decide which one to accept? Many people believe that laws apply not just in our local region but everywhere in spacetime. Are we justified in holding such beliefs given our finite and limited evidence? 
  \item \textit{The marks of the nomic:} There are certain features, such as simplicity, universality, exactness, and objectivity, that we normally associate with  laws (the nomic elements of reality). How should we understand those hallmarks, in light of the metaphysics and epistemology of laws? 
\end{enumerate}
Such questions do not have straightforward answers, and they cannot be directly tested in empirical experiments. They fall in the domain of philosophy. 

 The Great Divide in metaphysical debates about  laws is between Humeans, who think that laws are merely descriptions, and non-Humeans, who think that laws govern.\footnote{This is an oversimplification as there are some non-Humeans, such as Aristotelian Reductionists, who do not think that laws govern. See \S4.3.} Humeans maintain that laws merely describe how matter is distributed in the universe. In Lewis's version (\citeyear{LewisCounterfactuals, LewisNWTU, lewis1986philosophical}), laws are just certain efficient summaries of the distribution of matter in the universe, also known as the \textit{Humean mosaic}, an example of which is a four-dimensional spacetime occupied by particles and fields. All there is in fundamental reality is the Humean mosaic; nothing enforces patterns or moves particles or fields around. On the face of it, Humeanism is highly revisionary; it regards patterns in nature as ultimately unexplained.  A common theme in non-Humean views is that laws govern the distribution of matter. By appealing to the governing laws, the patterns are explained. How laws perform such a role is a matter of debate, and there are differences of opinion between reductionist non-Humeans such as \cite{ArmstrongWIALON} and primitivist non-Humeans such as \cite{MaudlinMWP}. 



Many physicists and philosophers have non-Humean intuitions. However, when they first encounter the philosophical literature on laws, they face a dilemma. They reject Humeanism, but they find traditional non-Humeanism unattractive. For example, some accounts explain laws in terms of other entities, such as Platonic universals or Aristotelian dispositions,  which are foreign to scientific practice. Other accounts severely limit the forms of laws one is allowed to consider. It is sometimes assumed that the governing view requires that laws must be \textit{dynamical} laws that \textit{produce} later states of the world from earlier ones, in accord with the direction of time that makes a fundamental distinction between past and future. Call this conception of governing \textit{dynamic production}. 
However, reflecting on the variety of kinds of laws that physicists present as fundamental, we find many that do not fit in the form of dynamical laws. These include principles of least action (that constrain physical history between two times), the Einstein equation of general relativity (which in its usual presentation is non-dynamical), and the Past Hypothesis (of a low-entropy boundary condition of the universe).  Moreover, even when physicists postulate dynamical laws,  dynamic production in accord with a fundamental direction of time does not seem essential to how these laws govern the world or explain the observed phenomena. Many physicists and philosophers regard the direction of time as an emergent feature of reality, not something put in by hand.
Hence, we have good reasons to consider more flexible and minimalist versions of non-Humeanism that better accommodate modern physics. 

I present and develop a minimal primitivist view (MinP) about laws of nature, introduced in   \citep{chenandgoldstein}, that disentangles the governing conception from dynamic production, and requires no reduction or analysis of laws into something else. It is a non-Humean view where laws govern the universe. On MinP:
\begin{itemize}
  \item (Primitivism) Fundamental laws are regarded as fundamental facts of the universe; they are not reducible or analyzable into universals, dispositions, or anything else. MinP regards laws as elements of fundamental reality. 
  \item (Minimalism) Fundamental laws govern by constraining the physical possibilities of the entire spacetime and its contents.\footnote{As a first approximation, I assume that spacetime is fundamental.  This assumption is not essential to MinP. One can consider non-spatio-temporal worlds governed by minimal primitivist laws. For those worlds, one can understand MinP as suggesting that laws constrain the physical possibilities of the world, whatever non-spatio-temporal structure it may have. Indeed, if one regards time itself as emergent,  one may find it natural to understand governing in an atemporal and direction-less sense. } They need not exclusively be dynamical laws, and their governance does not presuppose a fundamental direction of time.
\end{itemize}
MinP captures the essence of the governing view without taking on extraneous commitments. Because of the primitivism and the minimalism, MinP accommodates a variety of candidate fundamental laws. The flexibility of MinP is, I believe, a virtue. It is an empirical matter what forms the fundamental laws take on; one's metaphysical theory of laws should be open to accommodating the diverse kinds of laws entertained by physicists. MinP encourages openness. 

My goal is to introduce readers to some contemporary philosophical issues about laws of physics. I shall focus on MinP, as it provides a unified lens for thinking about such issues and a clear contrast from traditional accounts of laws. First, various conceptual connections are illustrated, more or less straightforwardly, by thinking about how laws constrain.   MinP is a useful entry point into this debate, as one can appreciate the core issues about laws without prior familiarity with deep issues in metaphysics (such as universals, dispositions, and Humean supervenience). 

Second, as MinP captures the essence of the governing view without taking on extraneous commitments, it provides the ideal non-Humean theory to contrast from Humeanism. We are able to better appreciate their fundamental difference, which is about explanatory priority. The minimalism of MinP also clarifies the epistemological puzzles about the discovery of physical laws. Our epistemic access to physical laws is based on certain epistemic principles regarding the simplicity and explanatory virtues of physical laws. It turns out that both Humeans and non-Humeans need to posit such epistemic principles  in addition to the metaphysical accounts about what laws are. 

Finally, the explicitness of  the epistemic principles allows us to re-examine our commitments to certain hallmarks of physical laws---simplicity, exactness, and objectivity. The metaphysical and epistemological discussions provide principled reasons for whether and how we should associate them with laws.

This volume is intended for advanced undergraduate students, graduate students, and professional researchers in philosophy of science, foundations of physics, and mathematical physics who are interested in the philosophical issues about laws.   I will keep technical details at an appropriate level, so that the volume is accessible to those who do not specialize in philosophy of physics. 

\section{Conceptual Connections}

Laws occupy a central place in a systematic philosophical account of the physical world.  What makes them interesting is their connections to a wide range of issues, such as ontology, modality, explanations, counterfactuals, causation, time, induction, determinism, chance, and fundamentality. 
(This section  can also serve as a standalone introduction to the conceptual foundations about physical laws that are often implicitly assumed in the philosophical literature.)

\subsection{Ontology, Nomology, and Possibility}


A well-formulated physical theory contains two parts: (1) a fundamental ontology about what things there are in the physical world,  and (2) a fundamental nomology about how such things behave. The two are deeply connected. We focus on (2) in the rest of this volume. Here we say a few words about (1) and how the two are connected. 

Let us start with a first-pass definition of the fundamental ontology of a theory. 
\begin{description}
  \item[Fundamental Ontology]  The fundamental ontology of a physical theory refers to the fundamental material objects, their fundamental properties,  and the spacetime they occupy, according to that theory. 
\end{description}
For a familiar example, consider a version of Newtonian gravitation theory. Its fundamental ontology has three components:
\begin{itemize}
  \item Fundamental material objects: $N$ particles
  \item Fundamental properties: their masses,  $(m_1, m_2, ..., m_N)$, and their trajectories in physical space, $(q_1(t), q_2(t), ..., q_N(t))$
  \item Spacetime:  $3$-dimensional Euclidean space, represented by the Cartesian coordinate space $\mathbb{R}^3$, and $1$-dimensional time, represented by $\mathbb{R}^1$
\end{itemize}
For simplicity, let us assume that all $N$ particles have equal mass $m=1$ in the chosen unit. We can define the following concepts: 


   \begin{figure}
\centering	
  \begin{subfigure}[b]{0.45\textwidth}
\tdplotsetmaincoords{60}{110}
\begin{tikzpicture}[scale=3, tdplot_main_coords]
    \coordinate (O) at (0,0,0);
    \draw[thick,->] (0,0,0) -- (1,0,0) node[anchor=north east]{$x$};
    \draw[thick,->] (0,0,0) -- (0,1,0) node[anchor=north west]{$y$};
    \draw[thick,->] (0,0,0) -- (0,0,1) node[anchor=south]{$z$};
    \tdplotsetcoord{P}{1}{30}{60}
    \filldraw [black] (1, 1,1) circle (0.5pt) node[right] {$q_2$};
    \filldraw [black] (0.5, 0.5,1) circle (0.5pt) node[right] {$q_1$};
  \end{tikzpicture}
  \caption{Physical space $\mathbb{R}^{3}$ }
  \end{subfigure}
    \begin{subfigure}[b]{0.45\textwidth}
\tdplotsetmaincoords{60}{110}
\begin{tikzpicture}
    \begin{axis}[
        xmin=0, xmax=10, 
        ymin=0, ymax=10,
        ytick=\empty,
        xtick=\empty,
        ]
    \end{axis}
        \filldraw [black] (1, 1) circle (1pt) node[right] {$Q = (q_1, q_2)$};
    \end{tikzpicture}
    \caption{State space $\mathbb{R}^{6}$}
  \end{subfigure}
  \caption{Configuration of a two-particle universe, represented in two spaces.}
\end{figure}
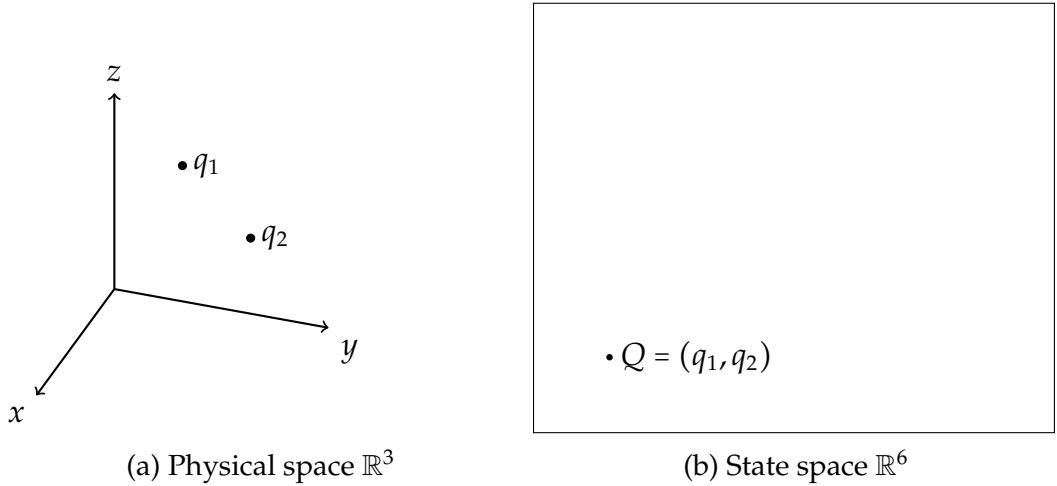

 \textit{(i) Physical states.} The fundamental physical state of the universe at time $t$ is the instantaneous state of the fundamental ontology at $t$, i.e. the arrangement of fundamental material objects and their properties at $t$.  In the example above, the state of the $N$-particle universe at time $t$ is a list $(q_1(t), q_2(t), ..., q_N(t))$ (see Figure 1a), together with the mass values that do not depend on time. Call $Q(t)=(q_1(t), q_2(t), ..., q_N(t))$  a \textit{configuration} of the universe.
 
 (It is often useful to consider other information, such as momenta of the $N$ particles, $(p_1(t), p_2(t), ..., p_N(t))$, alongside positions.  If we understand momenta as velocities (changes in positions) multiplied with mass, then momenta need not be fundamental properties of the particles. A state description with both positions and momenta,  $X(t) = (q_1(t), q_2(t), ..., q_N(t); p_1(t), p_2(t), ..., p_N(t))$, which includes more information than the fundamental physical state, can still be regarded as a physical state.)
 
  \textit{(ii) State spaces.} There are many possible states for the universe to be at any time. A space of all such possible states is a state space. The space of all possible configurations is called the \textit{configuration space}. Each point in the configuration space corresponds to a possible value of $Q(t)$, a possible list of the positions of $N$ particles in $\mathbb{R}^{3}$. The configuration space is represented by $\mathbb{R}^{3N}$ (Figure 1b). 
  
  (When it is useful to consider momenta in addition to particle positions, as in classical mechanics, we may define a space of higher dimensions called the \textit{phase space}. Each point in the phase space corresponds to $X(t)$, a possible list of the positions \textit{and momenta} of $N$ particles in $\mathbb{R}^{3}$. The list $X(t)$ is twice as long as $Q(t)$. The phase space is represented by $\mathbb{R}^{6N}$.)

  \textit{(iii) Physical histories.} We can consider a physical history of the $N$-particle universe in terms of physical states and state spaces. The most intuitive way is to represent the physical history as $N$ curves in physical space, corresponding to the positions of the $N$ particles at different times (Figure 2a). The state spaces provide mathematically convenient but more abstract representations. The physical history of the entire universe corresponds to a single curve in the high-dimensional configuration space, representing the configurations at different times (Figure 2b). (It also can be represented as a single curve in phase space.)\footnote{In the relativistic context, space and time become intertwined in such a way that there is no fundamental notion of physical state at a time. The more fundamental notion is the spacetime histories of $N$ particles. We may represent this as $N$ curves, also called \textit{world lines}, in a four-dimensional spacetime.} Notice that the concept of physical histories does not presuppose a direction of time. The arrow-less curves that represent physical histories can be regarded as direction-less histories that do not distinguish between the past and the future. The curves tell us whether event B is between events A and C, but not whether event B is earlier than C.

   \begin{figure}
\centering	
  \begin{subfigure}[b]{0.45\textwidth}
\tdplotsetmaincoords{60}{110}
\begin{tikzpicture}[scale=3, tdplot_main_coords]
    \coordinate (O) at (0,0,0);
    \draw[thick,->] (0,0,0) -- (1,0,0) node[anchor=north east]{$x$};
    \draw[thick,->] (0,0,0) -- (0,1,0) node[anchor=north west]{$y$};
    \draw[thick,->] (0,0,0) -- (0,0,1) node[anchor=south]{$z$};
    \tdplotsetcoord{P}{1}{30}{60}
      \draw[thick, black] (0.5, 0.5,1) node[below] {$q_1(t)$}  to [bend left=20] (1,1.6,2.2) ;
    \draw[thick, black] (1, 1,1) node[below] {$q_2(t)$}  to [bend left=16] (1,1.7,2) ;
  \end{tikzpicture}
  \caption{Physical space $\mathbb{R}^{3}$}
  \end{subfigure}
    \begin{subfigure}[b]{0.45\textwidth}
\tdplotsetmaincoords{60}{110}
\begin{tikzpicture}
    \begin{axis}[
        xmin=0, xmax=10, 
        ymin=0, ymax=10,
        ytick=\empty,
        xtick=\empty,
        ]
    \end{axis}
        \draw [thick, black] (1, 1) node[right] {$Q(t) = (q_1(t), q_2(t))$} to [bend left=35] (5.5, 4);
    \end{tikzpicture}
    \caption{State space $\mathbb{R}^{6}$}
  \end{subfigure}
  \caption{Physical history of a two-particle universe, represented in two spaces.}
\end{figure}
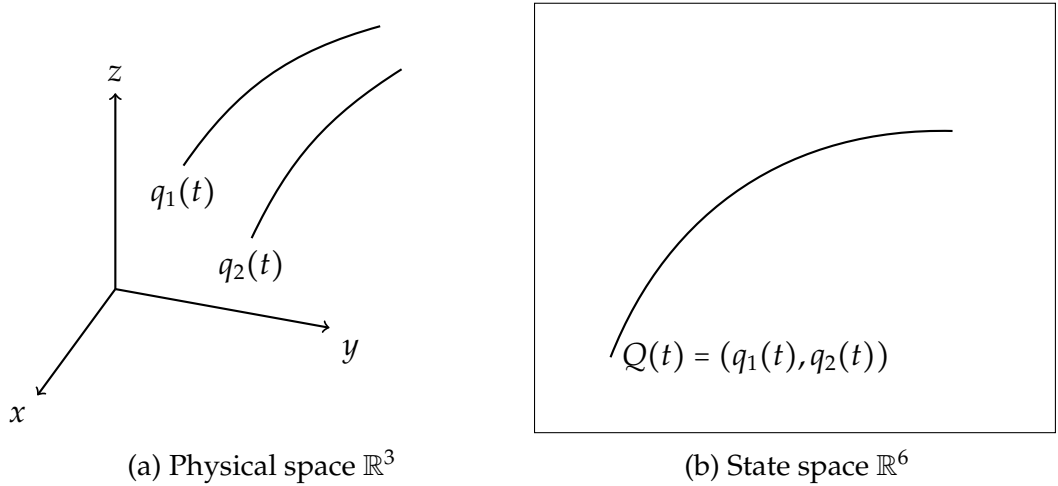

Next, let us define the fundamental nomology of a theory. 
\begin{description}
  \item[Fundamental Nomology] The fundamental nomology of a physical theory refers to the fundamental laws in the physical theory. 
\end{description}
The fundamental laws of Newtonian gravitation theory can be represented by the following equations:
\begin{itemize}
  \item The dynamical law:  $F=ma$, or equivalently $F_i(t) = m_i \frac{d^2 q_i(t)}{dt^2}$
  \item The force law: $F=GMm/r^2$, or equivalently $F_i(t) =\sum^N_{j \neq i} \frac{Gm_i m_j}{|q_i(t) - q_j(t)|^2}$, with $G$ the gravitational constant
\end{itemize}
For a Newtonian universe with $N$ particles, these laws tell us which set of physical histories are permissible. We may capture this with a natural interpretation in terms of possible worlds. A possible world is a logically consistent description of spacetime and its contents, namely the distribution of fundamental properties and material objects in spacetime. A possible world can be represented in multiple ways discussed above, with $N$ curves in physical space or spacetime, or a single curve in a high-dimensional state space.  The collection of all such worlds permitted by physical laws forms the set of nomological possibilities. 

More precisely, a nomologically possible world of theory $T$ is a logically consistent description of spacetime and its contents such that (1) the fundamental objects and properties are restricted to those kinds mentioned by the fundamental laws in $T$, and (2) their arrangement is compatible with those laws. In other words, a nomologically possible world of theory $T$ is a model of the laws of $T$. This definition can be specialized to the actual physical laws.  The actual world is a very special one---the spacetime with the actual arrangement of objects and their properties. We define the following: 




\begin{itemize}
  \item A possible world $w$: a spacetime and a distribution of material contents.\footnote{For simplicity, I assume that possible worlds have a fundamental spatio-temporal structure. This is important for defining determinism but not required for defining strong determinism, which suggests that the latter is more general than the former. I return to this point in \S2.4. }
  \item The actual world $\alpha$: the actual spacetime and the actual distribution of material contents.
  \item Material contents: material objects and their qualitative properties.
  \item $\Omega^T$: the set of possible worlds that satisfy the  fundamental laws  specified in theory $T$. 
  \item $\Omega_{\alpha}$: the set of  possible worlds that satisfy the actual fundamental laws of $\alpha$, i.e. the set of all nomologically possible worlds. 
\end{itemize}
Note that $\Omega_{\alpha} = \Omega^T$ only when $T$ is the actual theory of the world, i.e. the axioms of $T$ correspond to the fundamental laws governing $\alpha$.

The above definitions are global in character, as they concern entire possible worlds. Sometimes we are also interested in parts of worlds, such as whether certain events of a 30-minute interval is nomologically possible or impossible. We may define the following:

\begin{description}
  \item[Nomological Possibility] A sequence of events is nomologically possible if and only if it occurs in some nomologically possible world.
\end{description}

\begin{description}
  \item[Nomological Necessity] A sequence of events is nomologically necessary if and only if it occurs in every nomologically possible world.
\end{description}

   \begin{figure}
\centering	
  \begin{subfigure}[b]{0.45\textwidth}
\begin{tikzpicture}
    \begin{axis}[
        xmin=0, xmax=10, 
        ymin=0, ymax=10,
        ytick=\empty,
        xtick=\empty,
        ]
    \end{axis}
                         \draw [thick, black] (1,1) to [bend left=20]  coordinate[pos=0] (dl_j) (6,0.5); 
            \draw [thick, black] (1,2) to [bend left=20] coordinate[pos=0] (dl_j) (6,1.5);
        \draw [thick, black] (1,3) to [bend left=20] coordinate[pos=0] (dl_j) (6,2.5);
                \draw [thick, black] (1,4) to [bend left=20] coordinate[pos=0] (dl_j) (6,3.5);
                \draw [thick, black] (1,5) to [bend left=20] coordinate[pos=0] (dl_j) (6,4.5);
                     \draw [thick, black] (0.4, 5.4)  to [bend left=50] (6,0.2);
         \draw [thick, black] (0.4, 5)  to [bend left=50] (5.6,0.2);
                  \draw [thick, black] (0.4, 4.6)  to [bend left=50] (5.3,0.2);
                           \draw [thick, black] (0.4, 4.2)  to [bend left=50] (5,0.2);
        \draw [thick, black] (0.4, 3.8)  to [bend left=50] (4.5,0.2);
        \draw [thick, black] (1, 0.1)  to [bend left=40] (5, 5);
                \draw [thick, black] (0.1, 1)  to [bend left=50] (6, 5);
                          \draw [thick, black] (0.1, 2)  to [bend right=70] (6, 5);
        \draw [thick, black] (1, 1)  to [bend right=30] (6,3);
                \draw [thick, black] (0.2, 2)  to [bend left=30] (6,4);
        \draw [thick, black] (3, 5)  to [bend right=30] (3,0.1);
        \draw [thick, black] (4, 5)  to [bend right=30] (4,0.1);
        \draw [thick, black] (4, 5)  to [bend left=30] (5,0.1);
                \draw [thick, black] (0.1, 5)  to [bend left=40] (0.5,0.1);
    \end{tikzpicture}
  \caption{Arbitrary histories in state space}
  \end{subfigure}
    \begin{subfigure}[b]{0.45\textwidth}
\begin{tikzpicture}
    \begin{axis}[
        xmin=0, xmax=10, 
        ymin=0, ymax=10,
        ytick=\empty,
        xtick=\empty,
        ]
    \end{axis}
                   \draw [thick, black] (1,1) to [bend left=20]  coordinate[pos=0] (dl_j) (6,0.5); 
            \draw [thick, black] (1,2) to [bend left=20] coordinate[pos=0] (dl_j) (6,1.5);
        \draw [thick, black] (1,3) to [bend left=20] coordinate[pos=0] (dl_j) (6,2.5);
                \draw [thick, black] (1,4) to [bend left=20] coordinate[pos=0] (dl_j) (6,3.5);
                                     \draw [thick, black] (0.4, 5.4)  to [bend left=50] (6,0.2);
                           \draw [thick, black] (0.4, 4.2)  to [bend left=50] (5,0.2);
           \end{tikzpicture}
    \caption{Nomologically possible histories}
  \end{subfigure}
  \caption{Laws select a special subclass of physical histories as the nomologically possible histories.}
\end{figure}
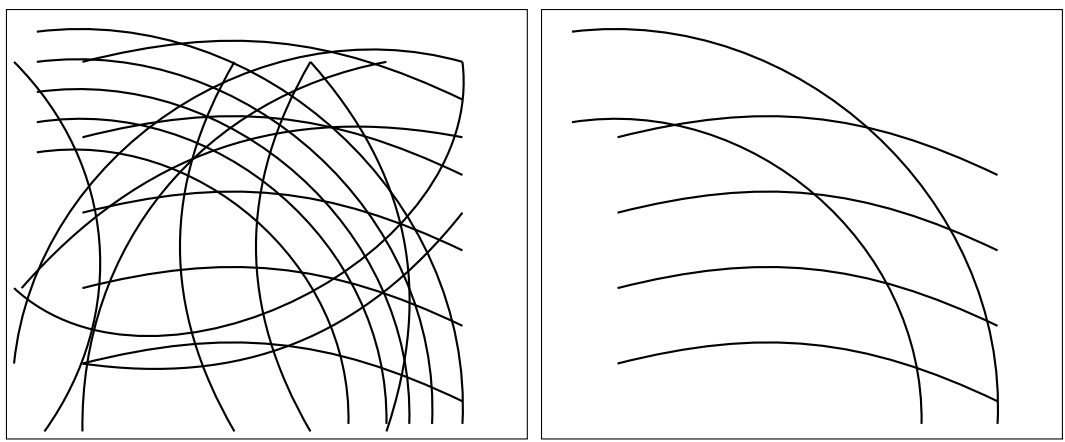

We may think of a choice of a fundamental ontology as pinning down abstract \textit{state spaces} that tell us what kind of physical states are available.  But it does not pin down which histories in those state spaces are nomologically possible (Figure 3a). The choice of a fundamental nomology selects a special subclass of histories as corresponding to the nomologically possible histories, which are also called physical possibilities (Figure 3b).

\subsection{Counterfactuals and Causation}

Laws \textit{support} counterfactuals. A counterfactual is a conditional of the form ``if A were the case, then B would be the case.'' (We often focus only on those for which A does not occur.) In a counterfactual, the consequence does not follow from the antecedent as a matter of logic; they are joined together by laws. For example, consider: 

\begin{description}
  \item[C1] If this match had been struck, it would have lit. 
\end{description}

\begin{description}
  \item[C2] If this match had been struck, it would not have lit. 
\end{description}

Suppose C1 is true and C2 false. For either one, the consequent and its negation are logically compatible with the antecedent. Hence, it is not logic alone that renders C1 true and C2 false. What non-logical fact is needed? Many agree that it involves some laws. But why laws in particular, but not just general facts of the form "every match that is struck in oxygen rich, dry, and no-wind condition is lit?" Generality is not sufficient, and lawfulness is crucial. Suppose every coin in my pocket is silver. Nevertheless, the following counterfactual is false: 

\begin{description}
  \item[C3] If this coin were in my pocket, it would have been silver. 
\end{description}
The problem is that the general fact \textit{every coin in my pocket is silver} is accidental. To support a counterfactual, the non-logical fact needs to have nomological necessity, corresponding to a law. To see this more clearly,  consider counterfactuals about physical systems:
 \begin{description}
  \item[C4] If this ice cube were placed in a cup of hot tea, it would have melted 30 seconds later. 
\end{description}

 \begin{description}
  \item[C5] If there were one more planet orbiting around the sun, it would have an elliptical orbit. 
\end{description}

 \begin{description}
  \item[C6] If the polarizer were oriented at 30 degrees from the median line, 25\% of the pairs of photons would have passed. 
\end{description}
To evaluate such counterfactuals, we need knowledge of the relevant laws (in thermodynamics, classical mechanics, and quantum mechanics). For C4, we can consider the nomologically possible worlds where  this ice cube were placed in a cup of hot tea, and check whether the ice cube is melted 30 seconds later in all (or most) of them. If the answer the yes, then C4 is true. For C5, we consider the nomologically possible worlds with a ninth planet orbiting around the sun and check whether it has an elliptical orbit. And so on. There are conceptual nuances and technical challenges in spelling out the exact nomic algorithms for evaluating such counterfactuals. (For a range of different proposals, see  \cite{LewisCDTA, albert2000time, albert2015after, kutach2002entropy} and \cite{LoewerCatSLaw}. For an updated discussion, see \cite{fernandes2023temporal}.)




Counterfactuals are related to deliberation, influence, and control. Rational deliberation depends on evaluating counterfactuals with various different suppositions, representing different options, and assessing their outcomes. If action A were selected, then outcome O would result.    If counterfactuals have nomic involvement, then so are those notions.  These further suggest the practical relevance of knowledge of counterfactuals and physical laws. 

Sometimes counterfactuals are also linked to causation. An influential approach seeks to analyze causation it in terms of counterfactual dependence. Roughly speaking, event A causes event E if and only if the following two counterfactuals are true: 

\begin{description}
  \item[C7] If A were the case, then C would be the case.
\end{description}

\begin{description}
  \item[C8] If A were not the case, then C would not be the case. 
\end{description}
For example, Suzy throws a rock at a window and the window breaks. Her throw causes the breaking of the window, because if she had not thrown the rock at the window it would not have broken. 
Due to various counterexamples to such an account, many people have given up the project of analyzing causation in terms of counterfactual dependence. Nevertheless, counterfactual dependence seems to capture an important aspect of causation. The central idea is also preserved in contemporary structural equation models of causation. For more on the latter, see \cite{sep-causal-models}. 


\subsection{Dynamic Production and the Direction of Time}

A concept closely related to causation is that of dynamic production. It is the idea that events in the past, together with the laws, bring about events in the future. (For a comparison of production and counterfactual dependence, see \cite{Hall2004two}.) For some people, dynamic production is constitutive of how laws govern and explain. Laws govern the universe by dynamically producing the subsequent states from earlier ones; an event is explained by appealing to the laws and the prior events that produce it. 


The emphasis on dynamic production is often associated with an emphasis on dynamical laws and the direction of time.  If dynamic production is how laws govern, perhaps laws should be dynamical laws that evolve the states of the universe successively in time. They should be exclusively what \cite{MaudlinMWP} calls \textit{Fundamental Laws of Temporal Evolution} (FLOTEs). Examples include Newton's $F=ma$, Schr\"odinger's equation, and Dirac's equation, but not Einstein equation, Gauss's law, or boundary-condition laws. Moreover, for dynamic production to make sense, the temporal development should be directed only from the past to the future. However, the laws in modern physics are blind to the past-future distinction; they are (essentially) time-reversal invariant in the sense that for any nomologically possible history going in one temporal direction, its temporal reverse is also nomologically possible. Where does the direction of time come from? A natural idea, on this picture, is to make the direction of time a fundamental feature of the universe. 

We may summarize this package of ideas as (1) a restriction of the form of laws: 
\begin{description}
  \item[Only FLOTEs]  The only kind of fundamental laws are fundamental laws of temporal evolution (FLOTEs). 
\end{description}
 (2) a commitment to dynamic production as how laws explain: 
\begin{description}
  \item[Dynamic Production]  Laws explain by producing later states of the universe from earlier ones. 
\end{description}
and (3) a metaphysical posit about the direction of time: 
\begin{description}
  \item[Temporal Direction Primitivism] The direction of time is a fundamental feature of the universe. 
\end{description}
Many people accept the package because it seems intuitive. Some build it into their theories of lawhood. An example is \cite{MaudlinMWP}, who expresses these ideas eloquently:  


\begin{quotation}
The universe started out in some particular initial state. The laws of temporal evolution operate, whether deterministically or stochastically, from that initial state to generate or produce later states. (p.174)
\end{quotation}

\begin{quotation}
This sort of explanation takes the term initial quite seriously: the initial state temporally precedes the explananda, which can be seen to arise from it (by means of the operation of the law). (p.176)
\end{quotation}

\begin{quotation}
The universe, as well as all the smaller parts of it, is made: it is an ongoing enterprise, generated from a beginning and guided towards its future by physical law. (p.182)
\end{quotation}

Despite the intuitive picture,  in my view dynamic production is inadequate for modern physics. It may be a useful heuristic picture to start out with, but once we see more examples of candidate laws and appreciate the explanations they provide, it is natural to replace the picture with something more flexible (allowing non-FLOTEs to be laws) and without a commitment to dynamic production or a fundamental direction of time.


An alternative approach, which I favor, is to understand the direction of time and dynamic production as important but derivative features of the physical world, partly explained by a boundary-condition law called the \textit{Past Hypothesis}. On this approach, the direction of time is understood in terms of an entropy gradient that arises from a new law---at one temporal boundary, the universe is in a low-entropy state. Given the Past Hypothesis as a nomic constraint, it is plausible to expect that most solutions to the dynamical equations will be ones that relaxes towards the thermodynamic equilibrium (maximum entropy) in the direction away from the temporal boundary where the Past Hypothesis applies. Hence, almost all the nomological possible worlds are such that they will display an entropy gradient, giving rise to an emergent  (non-fundamental) direction of time.  It has been argued that it is compatible with the Humean approach of laws, but it is also compatible with  non-Humeanism. On the non-Humean account I introduce in \S3, laws explain, not by producing the states of the universe in time, but constraining physical possibilities.  Dynamic production may also be regarded as a derivative concept.  

Allowing non-FLOTEs to be laws, the alternative approach opens up many new possibilities.  Still, we may sometimes prefer FLOTEs, but the preferences are not grounded in metaphysical prohibitions about the forms of laws, but in methodological and epistemic reasons that certain dynamical laws offer simple and compelling explanations of observed phenomena. As I shall argue, the alternative approach is better suited for accommodating the variety of kinds of laws in modern physics and understanding the explanations they provide.

\subsection{Determinism and Chance}

Determinism and indeterminism are properties of laws.  In his survey article, \cite{sep-determinism-causal} provides the following (first-pass) characterization of determinism (emphases original): 

\begin{description}
  \item[Determinism$_0$] The \textit{world} is \textit{governed by} (or is \textit{under the sway of}) determinism if and only if, given a specified \textit{way things are at a time t}, the way things go \textit{thereafter} is fixed as a matter of \textit{natural law}.
\end{description}
 As Hoefer notes, the word ``thereafter'' suggests that determinism in this sense is future-directed but not past-directed. 

 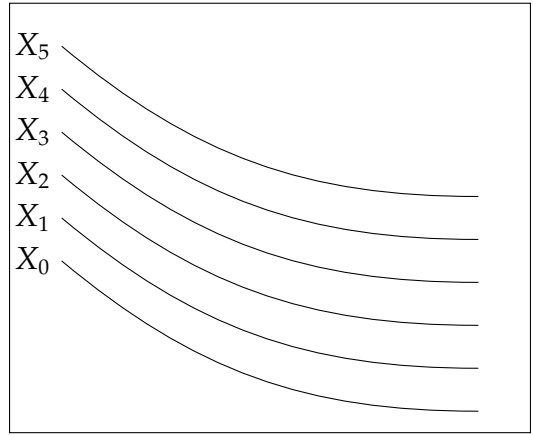
\begin{figure}
\centering	
\begin{tikzpicture}
    \begin{axis}[
        xmin=0, xmax=10, 
        ymin=0, ymax=10,
        ytick=\empty,
        xtick=\empty,
        ]
                    \draw (axis cs:1,4) to [bend right=20]  coordinate[pos=0] (dl_j) (axis cs:9,0.5); 
                            \fill (dl_j) circle (0pt) node[left] {$X_0$};
            \draw (axis cs:1,5) to [bend right=20] coordinate[pos=0] (dl_j) (axis cs:9,1.5);
                            \fill (dl_j) circle (0pt) node[left] {$X_1$};
        \draw (axis cs:1,6) to [bend right=20] coordinate[pos=0] (dl_j) (axis cs:9,2.5);
                                    \fill (dl_j) circle (0pt) node[left] {$X_2$};
                \draw (axis cs:1,7) to [bend right=20] coordinate[pos=0] (dl_j) (axis cs:9,3.5);
                            \fill (dl_j) circle (0pt) node[left] {$X_3$};
                \draw (axis cs:1,8) to [bend right=20] coordinate[pos=0] (dl_j) (axis cs:9,4.5);
                            \fill (dl_j) circle (0pt) node[left] {$X_4$};
                \draw (axis cs:1,9) to [bend right=20] coordinate[pos=0] (dl_j) (axis cs:9,5.5);
                            \fill (dl_j) circle (0pt) node[left] {$X_5$};
    \end{axis}
    \end{tikzpicture}
    \caption{Schematic illustration of a deterministic theory $T$.  $\Omega^T$ contains six nomologically possible worlds that do not cross in state space.}
\end{figure}

The core idea about determinism can be captured with nomological possibilities and without appealing to a direction of time.  Borrowing ideas from  \cite[pp.319-321]{MontagueFP},  \cite[p.360]{LewisNWTU}, and \cite[pp.12-13]{earman1986primer},  I define determinism as follows (also see Figure 4): 

\begin{description}
  \item[Determinism$_{T}$] Theory $T$ is  \textit{deterministic} just in case, for any two $w, w'\in \Omega^T$, if $w$ and $w'$ agree at any time, they agree at all times. 
\end{description} 
  
  \begin{description}
  \item[Determinism$_{\alpha}$]    The actual world $\alpha$ is \textit{deterministic} just in case, for any two $w, w'\in \Omega_{\alpha}$, if $w$ and $w'$ agree at any time, they agree at all times. 
\end{description}
Determinism is true just in case $\alpha$ is deterministic. My definitions correspond to what \cite[p.13]{earman1986primer} calls \textit{Laplacian determinism}.  The basic idea is that the nomologically possible worlds never cross in state space (like Figure 4 and unlike Figure 3b). By using four-dimensional spacetimes,  such definitions are more suitable for relativistic contexts as well as worlds without a fundamental direction of time.


Indeterminism is true just in case determinism is false. When a theory is indeterministic, it may also posit objective probabilities, some of which play roles similar to physical chances. There is an interesting question whether chance can co-exist with determinism. In classical and quantum statistical mechanics, even when the dynamical laws are deterministic, we may still posit probabilistic boundary conditions over nomologically possible initial states of the universe. \cite{ismael2009probability} has argued that they are indispensable for the predictive and explanatory success of those theories. \cite{barrett1995distribution} suggests that the probabilistic boundary conditions, in theories like Bohmian mechanics, are as important for the empirical adequacy of the theory as the dynamical equations.  For a survey about deterministic chances, see \cite[sect. 5]{sep-determinism-causal}.

 \begin{figure}
\centering	
\begin{tikzpicture}
    \begin{axis}[
        xmin=0, xmax=10, 
        ymin=0, ymax=10,
        ytick=\empty,
        xtick=\empty,
        ]
        \draw (axis cs:1,6) to [bend right=20] coordinate[pos=0] (dl_j) (axis cs:9,2.5);
                            \fill (dl_j)  node[left] {$X_0$};
        \end{axis}
    \end{tikzpicture}
    \caption{Schematic illustration of a strongly deterministic theory $T$. $\Omega^T$ contains exactly one nomologically possible world.}
\end{figure}
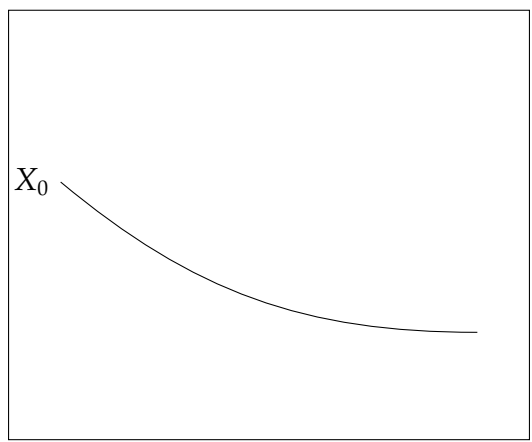

We can also define a stronger kind of determinism called \textit{strong determinism}.  According to \cite{roger1989emperor}, strong determinism is ``not just a matter of the future being determined by the past; the \textit{entire history of the universe is fixed}, according to some precise mathematical scheme, for all time'' (emphasis original, p.432). While Penrose defines strong determinism in terms of \textit{mathematical schemes}, I propose to define it in terms of \textit{fundamental laws}: a strongly deterministic theory of physics is one that, according to its fundamental laws, permits exactly one nomologically possible world; our world is strongly deterministic just in case it is the only nomologically possible world. I define it as follows (also see Figure 5): 

\begin{description}
	\item[Strong Determinism$_{T}$] Theory $T$ is strongly deterministic if $|\Omega^T| = 1$, i.e. its fundamental laws are compatible with exactly one possible world.
	\end{description}

\begin{description}
	\item[Strong Determinism$_{\alpha}$] The actual world $\alpha$ is strongly deterministic if $\Omega_{\alpha} = \{\alpha\}$. 
	\end{description}
	Under my definitions, strong determinism is stronger than determinism in a precise sense: whenever the definition of determinism is applicable, strong determinism logically implies determinism but not vice versa. 


Strong determinism is more general than determinism. There are circumstances where strong determinism applies but determinism does not (at least not naturally). That is because defining strong determinism only requires the minimal notion of the cardinality of the set of models, while defining determinism requires a notion of temporal agreement, which is not always guaranteed. For example, we can contemplate worlds without a fundamental spatio-temporal structure (such as those without metrical or topological time), of which there may not be natural extension of determinism. We may not be able to say whether such worlds are deterministic, for the concept simply may not apply. But even if $w$ is such a world, we can still assess the cardinality of $\Omega_w$, the set of models compatible with the fundamental laws that govern $w$.  $|\Omega_w|$ is either 1 or larger than 1. Hence, the earlier result is valid only when the proviso holds---whenever determinism is applicable.

There are many interesting works on free will, rational deliberation, and agency under determinism. For some examples, see \cite{hoefer2002freedom, ismael2016physics} and \cite{loewer2020mentaculus}. 
 Strong determinism has not received much attention in the philosophical literature until recently. See \cite{ChenCHETCV, chen2022strong} and \cite{adlam2021determinism}.  Several proposed theories of quantum cosmology aspire to be strongly determinism. See \cite{hartle1983wave} and \cite{page2009symmetric}.  Strong determinism is a possible feature of our physical theory, but \cite{hartle1996scientific, hartle1997quantum} goes further and suggests that it is a requirement.

 \subsection{Prediction and Explanation}

To make predictions about the future, or retrodictions about the past, we  rely on laws. For example, to predict the positions of planets in the solar system (relative to the sun), we need their current configuration plus laws of Newtonian gravitation theory to calculate their subsequent positions. We can also use the same laws to retrodict the positions of those planets in the last millions of years. Predictions and retrodictions are enabled by physical laws. 

Laws are also connected to scientific explanations. The search for physical laws is motivated by various ``Why'' questions. Why do apples fall at this rate from this height? Why do planets move in such orbits? Why do ice cubes melt in a cup of hot tea? Laws provide answers to such questions. 

 There is no sharp line between explanations and predictions (or retrodictions), as they often overlap. When there is an accurate prediction from the laws, there is often a corresponding explanation enabled by laws.  However, they can also come apart. Some laws provide accurate predictions but no satisfying explanation. For example, in orthodox quantum mechanics, we have a practical algorithm that makes successful predictions. The algorithm associates abstract mathematical objects on  Hilbert space to experimental setups and yields predictions about measurement outcomes.  Those postulates may be regarded as laws of orthodox quantum mechanics. They are, however, terrible candidates for laws, for they are disunified and not appropriately linked to a fundamental physical ontology. It would be far better to replace them with a simpler and more unified set of laws, explicitly defined over some fundamental physical states, from which those measurement postulates can be derived as theorems. Hence, solutions to the quantum measurement problem (see \cite{sep-qt-issues} and \cite{maudlin2019philosophy} for reviews) can be seen as providing better and more satisfying explanations. (Readers familiar with the recent discussion about `explainable' artificial intelligence (AI) may see a parallel.)

Let us connect prediction and explanation with the previous discussion about determinism and strong determinism. We can distinguish different kinds  of prediction and explanation. On determinism, we have conditional predictions: 

\begin{description}  
  \item[Conditional Prediction] Conditional on the state of the universe at some time (or states of the universe at some finite interval of time), one can in-principle deduce, using the fundamental laws, the state of the universe at any time. 
\end{description}
In contrast, strong determinism enables what I call \textit{strong prediction}:
\begin{description}  
  \item[Strong Prediction] One can in-principle deduce, using the fundamental laws alone, the state of the universe at any time. 
\end{description}

Similarly,  deterministic laws account for a general temporal pattern (cf. \cite{russell1913notion}): 
 \begin{description}
  \item[Conditional Explanation] Conditional on the state of the universe at some time  (or states of the universe at some finite interval of time), one can explain, using the fundamental laws, the state of the universe at any time. 
\end{description}
In contrast, strongly deterministic laws can explain more.  They underwrite conditional explanations such as the above but also account for unconditional facts such as: 
 \begin{description}
  \item[Strong Explanation] One can explain, using the fundamental laws alone, the state of the universe at any time.  
\end{description}

We may understand both types of explanations in the deductive-nomological (DN) model. On this approach, a scientific explanation contains two parts: an \textit{explanandum} describing the phenomenon to be explained, and an \textit{explanans} that account for the phenomenon. There are two requirements for a successful DN explanation. First, the explanation from the explanans to the explanandum should take the form of a deductive argument. Second, the explanans must involve, as premises for the deductive argument, at least one law of nature. As an example, consider the explanation of the orbit of Earth around the sun via Newtonian mechanics. The explanandum is the orbit. The explanans consist in facts about the mass of the Earth, the mass of the Sun, and Newton's law of universal gravitation and Newton's laws of motion. The exact orbit follows, as a mathematical solution, from the laws plus initial condition of the positions and velocities of massive objects. In the DN model, a conditional explanation is one that involves at least one law and at least one non-law as the explanans; a strong explanation is one where laws are the only explanans. 

Philosophers have raised many objections to the DN model as a universal model of scientific explanation \cite[sect. 2]{sep-scientific-explanation}.  However,  the DN Model is a simple and fruitful approach to think about explanations in physics. The difficulty of applying the DN model lies in the fact that we lack criteria to distinguish laws from non-laws. So the task of coming up with a satisfactory account of explanation in physics is bound up with the task of coming up with a satisfactory account of  laws. Insofar as simplicity is an important feature of  laws, it may be essential to explanations as well. How exactly simplicity is connected to laws, and how it figures in explanations, is a topic I shall return to in \S5.

\subsection{Problems of Induction}

Induction is the inference to unobserved phenomena from  observed ones.  It is essential to the scientific enterprise, but difficult to justify. There are two problems of induction widely discussed in the philosophical literature: Hume's problem and Goodman's problem (see \cite{sep-induction-problem} for a review). Both can be understood in relation to laws. 

Hume asks us to consider what justifies the inference from
\begin{description}
  \item[I1] All observed instances of bread of a particular appearance have been nourishing.
\end{description}
to
\begin{description}
  \item[I2] The next instance of bread of that appearance will be  nourishing.
\end{description}
We often appeal to a Uniformity Principle:
\begin{description}
  \item[UP] Nature is uniform.
\end{description}
This is intended to explain why events in nature continue the same way, why past and future are relevantly similar, and why the next instance of bread of that type will be nourishing. A major difficulty that Hume has identified is that we do not have non-circular justifications for inductive inference, because our justification for UP comes from induction. 

Many of us accept the rationality of induction. But given Hume's argument, it is difficult to provide a rational justification. One might take the lesson here to be that we should make a substantive assumption about the world as the basis for inductive inference. UP is a good candidate and should be taken as epistemically fundamental. 

However, Goodman's ``new riddle of induction'' shows that even UP is insufficient. Suppose, up to time $t$, all observed emeralds are green. Two interpretations of UP will give us divergent predictions. Under the assumption that, in nature, color distributions in terms of \textit{green} and \textit{blue} continue over time, then we infer that all emeralds are green and the emerald observed after $t$ will be green. However, we can also understand uniformity in terms of the temporal distribution of \textit{grue} and \textit{bleen}, where \textit{grue} means green and observed before $t$ and blue and observed at or after $t$. Under the alternative interpretation of UP, then, we should infer that all emeralds are grue and the emerald observed after $t$ will be blue.  This can also be generalized to more realistic predicates used in physics. 

Some philosophers have responded by suggesting we need a specific version of UP. It should license  normal inductive inferences but not ``gruesome'' ones. Consider: 
\begin{description}
  \item[UP+] Nature is uniform for phenomena described in natural predicates. 
\end{description}
On UP+, we accept the pattern of green and blue to continue over time. This supports the inductive generalization that future observed emeralds are green, and does not support the hypothesis that they are blue. 

However, there is something fundamentally wrong about UP and its more specific version.  Specified in fundamental / natural predicates, nature is not uniform; moreover, the uniformity of nature is not necessary for induction. 

For example, the mosaic we inhabit, described in terms of the matter distribution in spacetime, is manifestly non-uniform. We do not live in an empty universe that is completely homogeneous and isotropic,  exactly the same in all regions or directions. The spacetime region we occupy is quite different from regions with violent collisions of stars and merging of black holes.  What happens on earth is quite different even from a nearby patch---the core of the sun, where nuclear fusion converts hydrogen into helium. The variety and complexity in the matter distribution does not diminish our confidence in the viability and the success of induction.  In fact, non-uniformity of a certain kind is arguably necessary for the observed temporal asymmetries in our universe, which may be a precondition for induction.\footnote{See \cite{wallace2010gravity} and \cite{rovelli2019past} for the importance of the hydrogen-helium imbalance in the early universe to the existence of the relevant time asymmetries. }   

Instead of focusing on the mosaic, we should consider making a substantive assumption about  laws. Suppose we replace UP+ with the principle that we should expect that the law $L$ to be uniform. This is in the right direction, but it still has problems. Some people might understand the uniformity of $L$ to mean that it is of the form  ``for all $x$, if $Fx$ then $Gx$,'' which is a regularity, i.e. a universally quantified statement about the mosaic, holding for everything, everywhere, and everywhen. The problem is that the principle is vacuous, as any statement can be translated into a universally quantified sentence. That I have five coins in my pocket is equivalent to the statement that, for everything and everywhere and everywhen, I have five coins in my pocket. Suppose we understand the uniformity of $L$ to mean that it does not refer to any particular individual, location, or time. That version is too restrictive. There are candidate laws that do refer to particular facts, such as the Past Hypothesis of statistical mechanics, quantum equilibrium distribution in Bohmian mechanics, the Weyl curvature hypothesis in general relativity, and the No-Boundary Wave Function proposal in quantum cosmology (see \S3.3). These laws can be accepted on scientific and inductive grounds, and may be required to ultimately vindicate our inductive practice. Suppose we understand PU to mean that the same law applies to everywhere in spacetime. That version is again vacuous, as even an intuitively non-uniform law can be described by a uniform law with a temporal variation, such as 
 \begin{equation}\label{E1}
  F=ma \text{ for } (-\infty, t] \text{ and } F=(8m^9-\frac{1}{7}m^5+\pi m^3+km^2+m)a \text{ for } (t, \infty)
\end{equation}
The same law can be applied to everywhere in spacetime. 

 As I suggest in \S5, what induction ultimately requires is the reasonable simplicity of   laws (expressed in natural predicates and traded off with other explanatory virtues).

\subsection{Fundamentality}

The final issue  concerns fundamentality. In metaphysics, there is much discussion about fundamental ontology (e.g. particles, fields, and spacetime) and non-fundamental ontology (e.g. tables, bananas, and galaxies). There is also a distinction between fundamental laws and non-fundamental laws.   How should they be distinguished? 

Consider the following definition of fundamental laws: 
\begin{description}
  \item[Fundamental Laws] For any world $w$, fundamental laws of nature in $w$ are the non-mathematical axioms (basic postulates) of the  complete fundamental physical theory of $w$.
\end{description}
For a physical theory to be complete in world $w$, it needs to entail all the important regularities in $w$, including those described by  non-fundamental laws (such as laws of chemistry, biology, and so on). For a physical theory to be fundamental in $w$, it cannot be derived from another non-equivalent physical theory that is true in $w$.
Hence, in a quantum world, classical mechanics is not a fundamental theory, because it can be  derived  from quantum mechanics (via approximations in some limit). However, in a classical world, classical mechanics is a fundamental theory  because quantum mechanics is not true in such a world. The physical theory can employ some mathematics, but the mathematical axioms are not laws of nature. Hence, the fundamental laws of nature are the \textit{non-mathematical axioms}. 


Laws in $w$ are either fundamental  or non-fundamental in $w$. I require  non-fundamental laws in $w$ be derivable\footnote{The relevant derivations may also involve approximations and idealizations. } from  fundamental laws in $w$. But not all deductive consequences of fundamental laws are laws, for otherwise we could  trivialize the notion of laws by  using disjunction introduction.  Some deductive consequences  will be more important than others because they support counterfactuals and are extraordinarily useful and simple. Identifying the sufficient conditions for non-fundamental laws is an important project,  but I do not pursue it here. Instead, I suggest a necessary condition for a law to be non-fundamental in $w$: 
\begin{description} 
  \item[Necessary Condition for Non-Fundamental Laws] In any world $w$, if a law of nature is a non-fundamental law in $w$, then it can be (non-trivially) derived from the fundamental laws in $w$. 
\end{description}
Consequently, a law that cannot be so (non-trivially) derived in $w$ is a fundamental law in $w$. An example of a non-fundamental law in our world is the ideal gas law $PV=nRT$ that can be derived from the micro-physics.  Not all non-fundamental laws have been successfully derived from the fundamental axioms in physics, but what matters is that they can be.

In philosophy of physics, there are debates about the theoretical structure of quantum field theories, a theory that is remarkably accurate for certain predictions about the subatomic particles but employs methods and principles (such as the renormalization group) not straightforwardly interpretable as fundamental laws. Those debates may give us a richer understanding of how fundamental laws relate to non-fundamental laws. For recent discussions about these issues in relation to fundamentality, see \cite{mckenzie2022fundamentality} and \cite{williamsphilosophy}. 

What about fundamental laws themselves? Are they metaphysically fundamental? Or are they metaphysically explained by something else? We turn to these questions in the next section. 

\subsection{Summary}

Those conceptual connections are evidence of the centrality of laws in a comprehensive philosophy of science.  As a first step towards such a philosophy, one needs to develop a metaphysical account of what laws are.  

\section{Minimal Primitivism}

As an example of a metaphysical account of laws, I introduce the view about laws that Sheldon Goldstein and I call \textit{Minimal Primitivism} (MinP). 
  On MinP, laws are taken at face value. Fundamental  laws exist; they are fundamental features of reality, not reducible to anything else; we should simply start with fundamental laws and try to use them to illuminate other issues, such as those discussed in \S2. Laws govern the behavior of physical objects; they are not limited to FLOTEs. 

\subsection{Minimal Primitivism (MinP)}

According to MinP,  fundamental laws are ontological primitives that are metaphysically fundamental. They do not require anything else to exist. 
For  laws to govern, they are not required to dynamically produce or generate later states of the universe from earlier ones, nor are they required to presume a fundamental direction of time. On MinP, laws govern by constraining the physical possibilities (nomological possibilities). MinP is flexible regarding the form of laws. To summarize, the first part of the view is a metaphysical thesis: 

\begin{description}
  \item[Minimal Primitivism] Fundamental laws of nature are certain primitive facts about the world.  There is no restriction on the form of the fundamental laws.  They govern the behavior of material objects by constraining the physical possibilities. 
\end{description}

Even though there is no metaphysical restriction on the form of fundamental laws, it is rational to expect them to have certain nice features, such as simplicity and informativeness. On Humean Reductionism (\S4.1), those features are metaphysically constitutive of laws, but on MinP they are merely epistemic guides for discovering and evaluating the laws. At the end of the day, they are defeasible guides, and we can be wrong about the fundamental laws even if we are fully rational in scientific investigations. The second part of our view is  an epistemic thesis: 

\begin{description}
  \item[Epistemic Guides] Even though theoretical virtues such as simplicity, informativeness, fit, and degree of naturalness are not metaphysically constitutive of fundamental laws, they are good epistemic guides for discovering and evaluating them. 
\end{description}
where the theoretical virtues correspond to methodological principles that have been successfully applied in scientific practice. 

Let me offer some clarifications:

(i) \textit{Primitive Facts}. Fundamental laws of nature are certain primitive facts about the world, in the sense that they are not metaphysically dependent on, reducible to, or analyzable in terms of anything else. \x{If the concrete physical reality corresponds to a Humean mosaic, then fundamental laws are facts that transcend the mosaic. Many physicists may even regard fundamental laws as more important than the mosaic itself.}  Depending on one's metaphysical attitude towards mathematics and logic, there might be mathematical and logical facts that are also primitive in that sense. For example, arithmetical facts such as $2+3=5$ and the logical law of excluded middle may also be primitive facts that \x{transcend the concrete physical reality and} constrain the physical possibilities, since every physical possibility must conform to them. However, we do not think that fundamental laws of nature are purely mathematical or logical. Hence,  fundamental laws of nature are not those kinds of primitive facts. 

(ii)  \textit{The Governing Relation}.  On MinP,  laws govern by constraining the world (the entire spacetime and its contents). We may understand constraining as a \textit{primitive relation} between fundamental laws and the actual world. We can better understand constraining by drawing conceptual connections to physical possibilities. Laws constrain the world by limiting the physical possibilities and constraining the actual world to be one of them. In other words, the actual world is constrained to be compatible with the laws. To use an earlier example, $F=ma$ governs by constraining the physical possibilities to exactly those that are compatible with $F=ma$. If $F=ma$ is a  law that governs the actual world, then the actual world is a possibility compatible with $F=ma$. 

Constraint differs from dynamic production; it does not require a fundamental distinction between  past and  future, or one between earlier states and later states. What the laws constrain is the entire spacetime and its contents. In some cases, the constraint \x{imposed by a law} can be expressed in terms of differential equations that \x{may} be interpreted as determining future states from past ones. (But \textit{not all constraints need be like that}. We discuss some examples in \S3.2 and \S3.3.)

For a concrete example,  consider the Hamilton's equations of motion for $N$ point particles with Newtonian masses $(m_1, ..., m_N)$ moving in a 3-dimensional Euclidean space, whose positions and momenta are $(\boldsymbol{q_1}, ..., \boldsymbol{q_N}; \boldsymbol{p_1},...,\boldsymbol{p_N})$: 
\begin{equation}\label{HE}
\frac{d \boldsymbol{q_i}(t)}{d t} = \frac{\partial H}{\partial \boldsymbol{p_i}} \text{  ,  } \frac{d \boldsymbol{p_i}(t)}{d t} = - \frac{\partial H}{\partial \boldsymbol{q_i}}
\end{equation}
where $H = H(\boldsymbol{q_1}, ..., \boldsymbol{q_N}; \boldsymbol{p_1},...,\boldsymbol{p_N})$ is specified in accord with Newtonian gravitation:
\begin{equation}\label{H}
H = \sum^{N}_{i} \frac{\boldsymbol{p_i}^2}{2m_i} - \sum_{1\leq j < k\leq N} \frac{G m_j m_k}{|\boldsymbol{q_j} - \boldsymbol{q_k}|}
\end{equation}
Suppose equations (\ref{HE}) and (\ref{H}) are the fundamental laws that govern our world $\alpha$. 
Let $\Omega^H$ denote the set of solutions to (\ref{HE}) and (\ref{H}). Represented geometrically, $\Omega^H$ corresponds  to a special set of curves on the state space of the $N$-particle system---the phase space represented by $\mathbb{R}^{6N}$.
Saying that  (\ref{HE}) and (\ref{H}) govern our world implies that  $\alpha$ should be compatible with them. In other words, $ \Omega^H$ delineates the set of physical possibilities, and $\alpha \in \Omega^H$. 
 
 In this example, the dynamical equations are time-reversible. For every solution in $\Omega^H$, its  time reversal under $t \rightarrow -t$ and $\boldsymbol{p} \rightarrow - \boldsymbol{p}$ is also a solution in $\Omega^H$. Since the concept of governing in MinP does not presuppose a fundamental direction of time, two solutions that are time-reversal of each other \textit{can} be identified as the same physical possibility. (If one prefers the representation where the set of physical possibilities contains each possibility exactly once,  one can derive a quotient set $\Omega_{\alpha}^\ast$  from $\Omega_{\alpha}$ with the equivalence relation given by the time-reversal map.)

(iii) \textit{Nomic Equivalence}. We should not think that, in every case, a law is equivalent to the set of possibilities it generates. The two can be different. For example, there are many principles and equations that can give rise to the same set of possibilities denoted by $\Omega^H$. But we expect  laws to be simple. One way to pick out the set $\Omega^H$ is by giving a complete (and infinitely) long list of possible histories contained in $\Omega^H$. Another is by writing down simple equations, such as (\ref{HE}) and (\ref{H}), which express simple laws.  Hence, the equivalence of  laws is not just the equivalence of their classes of models. For two laws to be equivalent, it will require something more. 

It is an interesting and open question, on MinP, what more is required and how to understand the equivalence of  laws. It seems to me that their equivalence must be related to simplicity and explanations, because a central role for laws to play is to provide an illuminating account of natural phenomena (see \S3.2). A natural idea, then, is to say that nomic equivalence requires explanatory equivalence, for which simplicity is an important factor. Hence, two fundamental facts that differ in their relative complexity cannot express the same law. 
(For a survey of the related topic of theoretical equivalence, see \cite{weatherall2019part1, weatherall2019part2}.)

(iv) \textit{The Mystery Objection.} Some might object that our notion of governing is entirely mysterious \citep{BeebeeNGCLN}. The notion of governing seems derived from the notion of government and the notion of being governed. But laws of nature are obviously not imposed by human (or divine) agents. So isn't it mysterious that laws can govern? To that we reply that a better analogy for governing laws is not to human government, but to laws of mathematics and logic. Arithmetical truths such as $2+3=5$ and logical truths such as the law of excluded middle can also be said to constrain our world. That is, the actual world cannot be a world that violates those mathematical or logical truths. In fact, every possible world needs to respect those truths. In a similar way, laws of physics constrain our world. The actual world cannot be a world that violates the physical laws, and every physically possible world needs to respect those laws. Those modal claims reflect  physical laws and mathematical laws. We can also make sense of the difference in scope between those laws. Mathematical laws are more general than physical laws, in the sense that the former are compatible with ``more models'' than the latter.  In any case,  mathematical laws and logical laws can also be said to govern the universe in the sense of imposing formal constraints. They generate a class of models and constrain the actual world to be one among them. There is also a difference in epistemic access. In some sense, we discover mathematical and logical laws \textit{a priori}, without the need for experiments or observations, but we discover physical laws \textit{a posteriori}, empirically. 

We do not claim that the analogy with mathematical and logical laws completely eliminates the mystery of how physical laws govern. However, we think it dispels the objection as previously stated, in terms of how something can govern the world without being imposed by an agent. If there is more to the mystery objection, it needs to be stated differently. On MinP, laws govern by constraining, and constraining is what they do. This provides the oomph behind scientific explanations. (We return to this shortly.) However, \x{in contrast to other non-Humean accounts, such an oomph is minimalist}. It does not require dynamic production, and it does not require an extra process supplied by a mechanism or an agent. 

(v) \textit{Epistemic Guides}. On MinP, even though the Humean criteria for the best system are not metaphysically constitutive for lawhood, they are nonetheless excellent epistemic guides for discovering and evaluating them. 

Regarding Epistemic Guides, one might ask in virtue of what  those theoretical virtues are reliable guides for finding and evaluating laws. This is a subtle issue, one related to the problems of induction. Unlike  Humeans, we cannot appeal to a reductive analysis of laws. (Humeans face a similar issue, as their account raises the worry as to why the fundamental Humean mosaic is so nice that it can be summarized in a simple way after all. Humean supervenience does not by itself solve the problems of induction. See \S5.) We can offer an empirical justification: the scientific methodology works. In so far as those theoretical virtues are central to scientific methodology, they are good guides for discovering and evaluating laws, and we expect them to continue to work. Can they fail to deliver us the true laws? That is a possibility. However,  if the true fundamental laws are complicated and messy, scientists would not be inclined to call them laws. \x{Therefore, there are two aspects of Epistemic Guides: discoverability and believability. Whereas simplicity as a guide for discovering a law might raise the question as to why laws should be simple (so that simplicity would be a reliable guide), simplicity as a requirement for believability seems clearer: it may be the case that the law is not simple, but if it is not simple it will not be believable as the law.}  

(vi) \textit{Temporal Variations of the Laws.} According to MinP, can laws change with time? In particular, can fundamental  laws be time dependent in such a way that different cosmic epochs are governed by different laws? In principle, MinP allows that possibility. If there is scientific motivation to develop theories in which laws take on different forms at different times (or in different epochs),  that is sufficient reason to consider a set of laws that govern different times, or a single law that varies in form with time. As a toy example, if we have empirical or theoretical reasons to think that the laws of motion are different on the two sides of the Big Crunch, say Newtonian mechanics and Bohmian mechanics, then different sides of the Crunch can be governed by different laws, or by a single law with a temporal variation. Similarly, MinP is compatible with fundamental constants of nature that have spatial or temporal variations (\cite{dirac1937cosmological}; see \cite{uzan2011varying} for a review). Hence,  on MinP, time-translation invariance may fail even for fundamental laws. Why then should we expect laws to be ``uniform'' in time so that they are friendly to induction? This is an instance of the previous point (v), to which I return in \S5.  
   
 (vii)  \textit{Fundamental vs. Non-Fundamental Properties.}  According to MinP, can fundamental laws refer to non-fundamental properties, such as entropy or temperature? Many fundamental laws we put forward refer only to fundamental properties. But it is reasonable to consider candidate fundamental laws that refer to non-fundamental properties.  Epistemic Guides  allows for this, as long as the non-fundamental properties are not too unnatural (all things considered). In the case of the Past Hypothesis, for example, we may sacrifice  fundamentality of the property involved but gain a lot of informativeness and simplicity if we invoke the property of entropy. The version of the Past Hypothesis that refers to entropy can still govern by constraining the physical possibilities. (Another strategy is to revise our definition of fundamental property such that any property mentioned by a fundamental law is regarded as fundamental, although it may be analyzable in terms of other fundamental properties.  However, this may present a problem for certain views of fundamentality.)

  (viii) \textit{Fundamental vs. Non-Fundamental Laws.} According to MinP, how are fundamental laws distinguished from non-fundamental laws? MinP allows for a reductionist picture where non-fundamental laws, when properly understood, are reducible to fundamental laws. We can distinguish them in terms of derivability: non-fundamental laws can be (non-trivially) derived from fundamental laws (\S2.7). For example, the ideal gas law is less fundamental than Newton's laws of motion, in the sense that the ideal gas law can be derived from them in suitable regime. However, derivability may not be sufficient for non-fundamental \textit{lawhood}, as other factors, such as counterfactual and explanatory robustness, may also be relevant. 
    
    (ix) \textit{Related Views.}  \cite{adlam2021laws} independently proposes an account that is, in certain aspects, similar to MinP;  she also suggests we take seriously laws that do not have a time-evolution form. However, her account is not committed to primitivism and seems more at home in a structural realist framework.  \cite{meacham2023nomic}'s nomic-likelihood account is also similar to MinP but Meacham takes objective probabilities as the starting point.  \cite{carroll1994laws} is often called a primitivist about laws, though recently \cite{carroll2018becoming} distances his view from primitivism and suggests a non-Humean reductive analysis of laws in terms of causation.
    \cite{bhogal2017minimal} proposes a ``minimal anti-Humeanism'' on which laws are ungrounded (true) universal generalizations. It is compatible with primitivism, but it is less minimalist than MinP. For example, on Bhogal's view, laws cannot be singular facts about particular times or places. However, Bhogal (p.447, fn.1) seems open to relax the requirement that laws have to be universal generalizations. It would be interesting to see how to extend Bhogal's view  to do so. 

\subsection{Explanation by Simple Constraint}

On MinP, laws explain, but not by accounting for the \textit{dynamic production} of successive states of the universe from earlier ones. They explain by expressing a hidden simplicity, given by compelling constraints that lie beneath complex phenomena. A fundamental direction of time is not required for our notion of explanation.\footnote{This type of explanation, sometimes called ``constraint explanation,'' has been explored in the causation literature by \cite{ben2018causation} and non-causal explanation literature by \cite{lange2016because}. Their accounts, with suitable modifications, may apply here. \x{See \cite{hildebrand2013can} for a critical discussion of primitive laws and explanations.} His criticisms to primitivism are addressed by the introduction of Epistemic Guides on MinP. }

In a world governed by Newtonian mechanics, particles travel along often complicated trajectories because that is implied by the simple fundamental law $F=ma$. Laws explain only when they can be expressed by simple principles or differential equations. It is often the case  that the complicated patterns we see in spacetime can be derived from  simple rules that we call laws. 

Fundamental laws need not be time-directed or time-dependent. They may govern purely spatial distribution of matter. For example,  Gauss's law

\begin{equation}\label{Gauss}
\nabla \cdot \textbf{E} = \rho
\end{equation}
 in classical electrodynamics---one of Maxwell's equations---governs the distribution of electric charges and the electric field in space.

Often the explanation that laws provide involves deriving striking,  novel, and unexpected patterns from  simple laws. The relative contrast between the simplicity of the law and the complexity and richness of the patterns may indicate that the law is the correct explanation of the patterns.  

 For a toy example, consider the Mandelbrot set in the complex plane (Figure 6), produced by the simple rule that a complex number $c$ is in the set just in case the function 

\begin{figure}
\centerline{\includegraphics[scale=0.4]{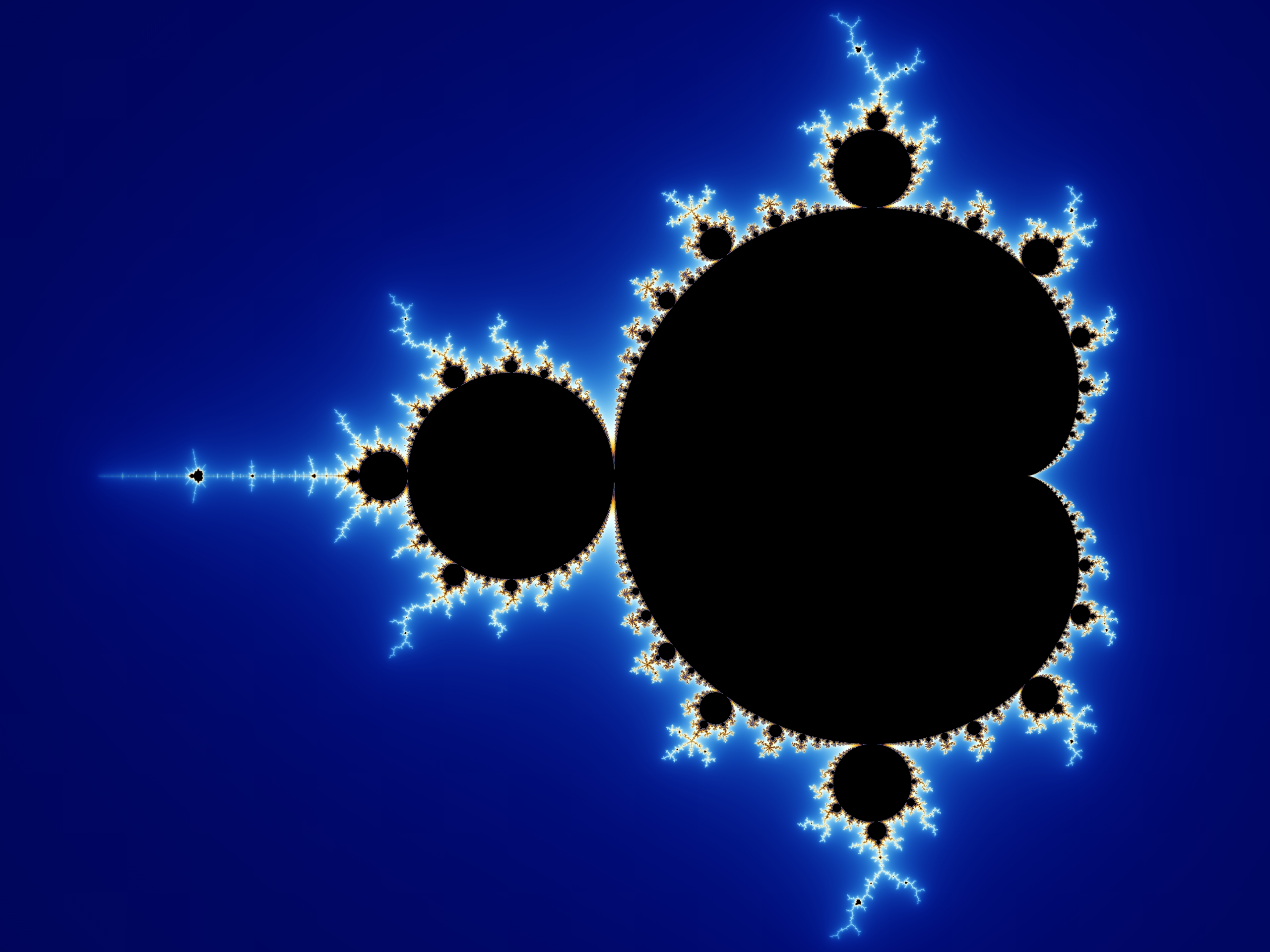}}
\caption{The Mandelbrot set with continuously colored environment. Picture created by Wolfgang Beyer with the program Ultra Fractal 3, CC BY-SA 3.0, https://creativecommons.org/licenses/by-sa/3.0, via Wikimedia Commons}
\end{figure}

\begin{equation}\label{Mandel}
	f_c (z) = z^2 +c 
\end{equation}
does not diverge when iterated starting from $z=0$. (For example, $c=-1$ is in this set but $c=1$ is not, since the sequence $(0,-1,0,-1,0,-1, ...)$ is bounded but $(0,1,2,5,26,677,458330, ...)$ is not.  For a nice description and visualization, see \cite[ch.4]{roger1989emperor}.) Here, a relatively simple rule yields a surprisingly intricate and rich pattern in the complex plane, a striking example of what is called the fractal structure. Now regard the Mandelbrot set as corresponding to the distribution of matter over (a two-dimensional) spacetime, the fundamental law for the world might be the rule just described. 
What is relevant here is that given just the pattern we may not expect it to be generated by any simple rule. It would be a profound discovery in that world to learn that its complicated structure is generated by the aforementioned rule based on the very simple function 	$f_c (z) = z^2 +c $. On our conception, it would be permissible to claim that the simple rule expresses the fundamental  law, even though it is not a law for dynamic production.  The Mandelbrot world is also an example of strong determinism. 

The previous examples illustrate some features of explanation on MinP:
\begin{enumerate}
  \item Laws explain by constraining the physical possibilities in an illuminating manner. 
  \item Nomic explanations (explanations given by laws) need not be dynamic explanations; indeed, they need not involve time at all. 
  \item Explanation by striking constraint can be especially illuminating when  an intricate and rich pattern can be derived from a simple rule that expresses the constraint imposed by a law.  
\end{enumerate}


On MinP, more generally, there are two ingredients of a successful scientific explanation: a metaphysical element and an epistemic one. It must refer to the objective structure in the world, but it also must relate to our mind, remove puzzlement, and provide an understanding of nature. We suggest that a successful scientific explanation that fundamental laws provides should contain two aspects: (i) metaphysical fundamentality and (ii) simplicity. 

The first aspect concerns the metaphysical status of fundamental laws: they should not be mere summaries of, or supervenient on, what actually happens; moreover, \x{what the laws are} should not depend on our actual practice or beliefs. This aspect is the \textit{precondition} for having a non-Humean account of scientific explanations. On MinP, the precondition is fulfilled by postulating fundamental laws as primitive (metaphysically fundamental) facts that constrain the world. The constraint provides the needed \textit{oomph} behind scientific explanations. Here lies the main difference between MinP and Humean Reductionism. 

The second aspect concerns how fundamental laws relate to us. Constraints, in and of themselves, do not always provide satisfying explanations.  Many constraints are complicated and thus insufficient for understanding nature.  What we look for in the final theory of physics is not just any constraint but simple, compelling ones that ground observed complexities of an often bewildering variety. The explanation they provide corresponds to an insight or realization that leads us to say, ``Aha! Now I understand.''  Often, simplicity is  related to elegance or beauty. As Penrose reminds us:
\begin{quotation}
  Elegance and simplicity are certainly things that go very much together. But nevertheless it cannot be quite the whole story. I think perhaps one should say it has to do with \textit{unexpected} simplicity, where one imagines that things are going to be complicated but suddenly they turn out to be very much simpler than expected. It is not unnatural that this should be pleasing to the mind. \cite[p.268]{penrose1974role}
\end{quotation}
The sense of unexpected simplicity is illustrated in the toy example of the Mandelbrot set as well as the  laws discovered by Newton, Schr\"odinger, and Einstein. 

Moreover, the second aspect of scientific explanation illuminates our principle of Epistemic Guides.  It is obvious that fundamental laws should be empirically adequate and consistent with all phenomena. But why should we expect them to be simple? On our view, it can partly be answered by thinking about the nature of scientific explanations. If successful scientific explanations require simple laws, then laws should be sufficiently simple to perform the explanatory role. One might press further and ask why laws should perform such roles and why scientific explanations can be successful. But they can be raised \textit{on any  account of laws} (see \S5). This is related to Hume's problem of induction \citep{sep-induction-problem}, which may be regarded as the problem of giving a non-circular justification for the uniformity of nature. Indeed, if we accept MinP and that the fundamental laws are simple, we have eliminated the problem of induction. If laws are simple, they may be completely uniform in space and time or else provide a simple rule that specifies how they change over space or time (\S3.1.vi). Thus,  the problem of induction reduces to the problem of justifying our acceptance of simple laws. In this sense, the problem of induction is not an additional problem over and above the justification of Epistemic Guides.

\subsection{Examples and Further Clarifications}

To further clarify MinP, I discuss some examples of dynamical laws and non-dynamical constraint laws. There is no difficulty accommodating them, as they can be understood as laws that constrain physical possibilities. Laws that involve intrinsic randomness present an interesting challenge. 

%

\subsubsection{Dynamical Laws}
Let us take a dynamical law to be any law that determines how objects move or things change. Sometimes the label is restricted to dynamical laws that can be understood as FLOTEs that guide the development of the universe in time. 

\textit{Hamilton's Equations.} Consider  classical mechanics for $N$ particles, described by Hamilton's equations of motion (\ref{HE})  with a Hamiltonian specified in (\ref{H}), \x{a  paradigmatic example of a FLOTE.} Hamilton's equations are differential equations of a particular type: they admit initial value formulations.  An intuitive way of thinking about dynamical laws is to understand them as evolving the initial state of the world into later ones.  However, this view is not entirely natural  for such a  system. The view requires momenta to be part of the intrinsic state of the world at a time; but it seems more natural to regard them as aspects of extended trajectories,  spanning  continuous intervals of time.  Regarding governing as dynamic production leads to  awkward questions about instantaneous states and whether they include velocities and momenta. 

 The situation becomes even more complicated with relativistic spacetimes having no preferred foliation of equal-time hypersurfaces. If there is no objective fact about which events are simultaneous, there is no unique prior Cauchy surface that is responsible for the production of any later state. This seems to detract from the intuitive idea of dynamic production as a relation with an objective input, making it less natural in a relativistic setting.\footnote{Christopher Dorst raised a similar point in personal communication. See also \cite{dorstproductive}.}

Instead of demanding that laws govern by producing subsequent states from earlier ones, we can regard laws as constraining the physical possibilities of spacetime and its contents. There is no difficulty accommodating the above example or any other type of dynamical laws. A dynamical law specifies a set of histories of the system and need not be interpreted as presupposing a fundamental direction of time. The histories the laws allow can often be understood as direction-less histories, descriptions of which events are temporally between which other events. 

A dynamical law such as (\ref{HE}) governs the actual world by constraining its history to be one allowed by (\ref{HE}). And MinP requires no privileged splitting of spacetime into space and time, as the physical possibilities can be stated in a completely coordinate-free way in terms of the contents of the 4-dimensional spacetime. 

\textit{Principles of Least Action.} Besides dynamical laws of Hamiltonian form, other kinds of equations and principles are often employed even for Hamiltonian systems. Consider, for example, Hamilton's principle of least action: this requires that  for a system of $N$ particles with Cartesian coordinates $q=(\boldsymbol{q_1}, \boldsymbol{q_2}, ..., \boldsymbol{q_N})$: 

\begin{equation}\label{PLA}
  \delta S = 0 
\end{equation}
where $S=\int^{t_2}_{t_1} L(q(t), \dot{q}, t) dt$, with   $\dot{q}=q(t)/dt$,  $\delta$  the first-order variation of $S$ corresponding to small variation in $q(t)$ with $q(t_1)$ and $q(t_2)$ fixed, and $L$,  the Lagrangian, is the kinetic energy minus the potential energy of the system of $N$ particles. While mathematically equivalent to Hamilton's equations, the principle of least action feels very different from a law expressing dynamic production. 
 For those who take dynamic production to be constitutive of governing, the principle of least action cannot be the fundamental governing law. They would presumably need to insist that the universe is genuinely governed by some law of a form such as (\ref{HE}), with the principle of least action arising as a theorem. On MinP, there is no problem regarding the principle of least action as a candidate fundamental law, with no need for it to be derived from anything else.  For a universe to obey the principle, its history must be one compatible with (\ref{PLA}). That is the sense in which it would govern our universe.

\textit{Wheeler-Feynman Electrodynamics.} Physicists have also considered dynamical equations that cannot be reformulated in Hamiltonian form. On MinP, there is no prohibition against laws expressed by such equations. For example,  \cite{wheeler1945interaction, wheeler1949classical} considered equations of motion for charged particles that involve both retarded fields ($F_{ret}$) and advanced ones ($F_{adv}$). On their theory, the trajectory of a charged particle depends  on charge distributions in the past (corresponding to $F_{ret}$) as well as  those in the future (corresponding to $F_{adv}$). Since the total field acting on particle $j$ is $F_{tot} = \sum_{k \neq j} \frac{1}{2}(^{(k)}F_{ret}+{}^{(k)}F_{adv})$, the equation of motion for particle $j$ of mass $m_j$, charge $e_j$, and spacetime location $q_j$ is
\begin{equation}
	m_j \ddot{q}_j^{\mu} = e_{j}\sum_{k \neq j} \frac{1}{2}(^{(k)}F_{ret}^{\mu \nu} +  {}^{(k)}F_{adv}^{\mu \nu})  \dot{q}_{j,\nu}
\end{equation}
with  the dot the time derivative with respect to proper time,  $^{(k)}F_{ret}$ the \x{retarded field contributed by the past trajectory of particle $k$, and $^{(k)}F_{adv}$ the advanced one, involving the future trajectory of particle $k$}. (For more details, see \cite{deckert2010electrodynamic} and \cite{lazarovici2018against}.) It is unclear how to understand the above equation in terms of dynamic production.  In contrast, it is clear on MinP: the fundamental law corresponding to such equations can be regarded as \x{imposing} a constraint on all trajectories of charged particles in spacetime.



\textit{Retrocausal Quantum Mechanics.}  There have been proposed reformulations of quantum mechanics that involve two independent wave functions of the universe: $\Psi_i(t)$ evolving from the past and $\Psi_f(t)$ evolving from the future. Some such proposals, motivated by a desire to evade  no-go theorems or preserve time-symmetry,  implement retrocausality or backward-in-time causal influences  \citep{sep-qm-retrocausality}. Consider \cite{sutherland2008causally}'s causally symmetric Bohm model, which specifies an equation of motion governing $N$ particles moving in a 3-dimensional space under the influence of both $\Psi_i(t)$ and $\Psi_f(t)$:

\begin{equation}
	\frac{d\boldsymbol{Q_j}(t)}{dt} = \frac{Re(\frac{\hbar}{2im_ja} \Psi_f^\ast \nabla_j \Psi_i)}{Re(\frac{1}{a}\Psi_f^\ast \Psi_i) } (Q (t), t)
\end{equation}  
with $Q(t)= (\boldsymbol{Q_1}(t), ... , \boldsymbol{Q_N}(t)) \in \mathbb{R}^{3N}$  the configuration of the $N$ particles at time $t$,  $m_j$ the  mass of particle $j$, and \x{$a= \int \Psi_f^\ast(q,t)\Psi_i(q,t)dq$}. It is unclear whether Sutherland's theory is viable; it also has many strange consequences. Nevertheless,  MinP is compatible with regarding the above equation \x{as expressing a fundamental law that constrains particle trajectories in spacetime} (even though we may have other reasons to not endorse the theory).  


\textit{The Einstein Equation.}  In general relativity, the fundamental equation is the Einstein equation: 
\begin{equation}\label{EFE}
  R_{\mu\nu} - \frac{1}{2} R g_{\mu \nu} = k_0 T_{\mu \nu} + \Lambda g_{\mu \nu}
\end{equation}
where $R_{\mu\nu}$ is the Ricci tensor, $R$ is the Ricci scalar, $g_{\mu\nu}$ is the metric tensor, $T_{\mu\nu}$ is the stress-energy tensor, $\Lambda$ is the cosmological constant, $k_0 = 8\pi G/c^4$ with $G$  Newton's gravitational constant and $c$  the speed of light. Roughly speaking, the Einstein equation is a constraint on the relation between the geometry of spacetime and the distribution of matter (matter-energy) in spacetime.  On MinP, we have no problem taking the equation itself as expressing a fundamental law of nature, one that constrains the actual spacetime and its contents. \x{If equation (\ref{EFE}) governs our world in the sense of MinP, then (\ref{EFE})} expresses a fundamental fact that does not supervene on or reduce to the actual spacetime and its contents. 

There are ways of converting equation (\ref{EFE}) into FLOTEs that are suitable for a dynamic productive interpretation.  If certain constraints are met, the Einstein equation can be decomposed in such a way that evolves an ``initial data'' of a 3-dimensional hypersurface forward in time. 
A famous example is the ADM formalism \citep{arnowitt62}.  However, they often discard certain solutions (such as spacetimes that are not globally hyperbolic). 
For non-Humeans who take dynamic production as constitutive for governing or explanations, those reformulations will be necessary. For them, the true laws of spacetime geometry should presumably be expressed by equations that describe the evolution of a 3-geometry in time.  In contrast, on MinP there is no metaphysical problem for taking the original Einstein equation as a fundamental  law.  The Einstein equation is simple and elegant and is generally regarded as the fundamental law in general relativity. We prefer not to discard or modify it on metaphysical grounds.

The Einstein equation allows some peculiar solutions. A particularly striking class of examples are spacetimes with closed timelike curves (CTCs). For MinP, there would seem to be no fundamental reason why such a possibility should be precluded. 
But the possibility of CTCs is precluded if we insist on dynamic production, since CTCs may lead to an event that dynamically produces itself. 

\subsubsection{Non-Dynamical Constraint Laws}

The examples mentioned earlier  are explicitly related to development in time. There are also important equations and principles that are not. We call them non-dynamical constraint laws. The minimal notion of governing easily applies to them.  For example, some purely spatial constraints on the universe may be thought of as  laws.  In \S3.2 we considered two examples of such laws---(\ref{Gauss}) and (\ref{Mandel}). Here we consider some more. 

\textit{The Past Hypothesis.} In the foundations of statistical mechanics and thermodynamics, followers of Boltzmann have proposed a candidate fundamental law of physics that Albert (2010) calls the Past Hypothesis (PH). It is a special boundary condition that is postulated to explain the emergent asymmetries of time in our universe, such as the Second Law of Thermodynamics. Here is one way to state it: 

\begin{description}
  \item[PH]  At one temporal boundary of the universe, the universe is in a low-entropy state.
\end{description}
This statement of PH is vague. We may be able to make it more precise by specifying the low-entropy state in terms of the thermoydnamic properties of the universe or in terms of some geometrical properties \citep{penrose1979singularities}. Penrose's version in general relativity, called the Weyl curvature hypothesis (WCH), renders it as follows: 

\begin{description}
  \item[WCH] The Weyl curvature $C_{abcd}$ vanishes at any ``initial'' singularity. 
\end{description}
(For an extension of this idea to loop quantum cosmology, see \cite{ashtekar2016initial, ashtekar2016quantum}.)
Let us use $\Omega_{PH}$ to denote the set of worlds compatible with PH.  If it is plausible that PH is a candidate fundamental law \citep{chen2020harvard}, then the metaphysical account of laws should make room for a boundary condition to be a fundamental law. On MinP, such an account is no problem.  Together, PH and dynamical laws can govern the actual world by constraining it to be one among the histories compatible with all of them. They require the actual world (history) lie in the intersection  $\Omega^{PH} \cap \Omega^{DL}$, where the latter denotes the set of histories compatible with the dynamical laws. However, PH is not a governing law in the sense of dynamic production. 

 
 \textit{The Initial Projection Hypothesis.} In a quantum universe, we have theoretical resources to postulate a stronger version of PH where the the law selects a particular initial microstate, represented by a mixed-state density matrix. In \cite{chen2018IPH}, I call this the Initial Projection Hypothesis (IPH):
 \begin{description}
  \item[IPH]  At one temporal boundary of the universe, the universal quantum state is the normalized projection onto the Past-Hypothesis subspace.
\end{description}
 This version of the boundary-condition law can be exact without incurring a theoretical cost called untraceability \citep{chen2018NV}. Since it pins down a unique initial microstate (in terms of a fundamental density matrix), together with the deterministic evolution equation for the quantum state, IPH yields a strongly deterministic theory. 
 
  \textit{The No-Boundary Proposal.} A famous initial condition for a quantum universe, proposed by two pioneers of quantum cosmology \cite{hartle1983wave}, is the idea that the universe has no temporal boundaries; the spacetime geometry smoothly rounds off and shrinks to a point in the ``past.'' They call it the No-Boundary Wave Function (NBWF):
   \begin{description}
  \item[NBWF]  The universal quantum state satisfies the Hartle-Hawking formula (\citeyear{hartle1983wave}), calculated over certain spacetime geometries that  shrink to a point in the ``past.''  
\end{description}
 NBWF was proposed to solve the problem that quantum cosmology does not yield any prediction unless one posits a boundary condition. \cite{hartle1996scientific, hartle1997quantum} regards this as a fundamental law of the universe on par with the dynamic laws. Together with deterministic dynamics, if the Hartle-Hawking formula has a unique solution, the theory will also be strongly deterministic.

\textit{Conservation Laws and Symmetry Principles.} According to a traditional perspective, symmetries such as those of rotation, spatial translation, and time translation are properties of the specific equations of motion. By Noether's theorem, those symmetries yield various conservation laws as theorems rather than postulates that need to be put in by hand. On that perspective, symmetries and conservation laws can be regarded as ontologically derivative of the fundamental laws, and are compatible with all metaphysical views on laws. 

According to a more recent perspective, symmetries are fundamental. See for example:  \cite{wigner1964symmetry, Wigner1985} and \cite{steven1992dreams}. \cite{lange2009laws} calls them \textit{metalaws}. For example, Wigner describes symmetries as ``laws which the laws of nature have to obey'' \cite[p.700]{Wigner1985}  and suggest that ``there is a great similarity between the relation of the laws of nature to the events on one hand, and the relation of symmetry principles to the laws of nature on the other'' \cite[p.957]{wigner1964symmetry}.  I do not take a firm stance on this perspective. Nevertheless, the perspective is compatible with MinP. If there is a symmetry principle $K$ that a fundamental law of nature $L$ must obey, then both $K$ and $L$ are fundamental facts, where $K$ constrains $L$ in the sense that the physical possibilities generated by $L$ are invariant under the symmetry principle $K$, and any other possible fundamental laws are also constrained by $K$. This introduces further ``modal'' relations in the fundamental facts beyond just the constraining of the spacetime and its contents by $L$.


Methodologically, one might prefer theories with dynamical laws, especially FLOTEs, to those without them. MinP allows this preference. Even though MinP does not restrict laws to FLOTEs,  the principle of Epistemic Guides suggests that we look for simple and informative laws.   FLOTEs, when they admit initial value formulations, may come with such theoretical virtues  \citep[ch.7-8]{callender2017makes}. The preference for FLOTEs and dynamical laws more generally may be explained by a preference for laws that strike a good balance between simplicity and informativeness. 

\subsubsection{Probabilistic Laws}

Candidate fundamental physical theories can also employ probability measures and distributions. Such measures and distributions can be objective, and they may be called objective probabilities. The probabilistic postulates in physical theories may well be lawlike, even though the nature of those probabilities is a controversial issue. 

 There are two types of probabilistic postulates in physics: (i) stochastic dynamics and (ii) probabilistic boundary conditions. We  start with (i) as it is more familiar. 
Consider the Ghirardi-Rimini-Weber theory  in quantum mechanics, a theory in which observers and measurements do not have a central place and in which the quantum wave function spontaneously collapses  according to precise probabilistic rules. On the GRW theory,  the wave function of the universe $\Psi(t)$ evolves unitarily according to the Schr\"odinger equation but is interrupted by random collapses. The probabilities of where and when the collapses occur are fixed by the theory.  
(For details, see \cite{ghirardi1986unified} and \cite{sep-qm-collapse}.) 

For an example of (ii), consider Albert and Loewer's Mentaculus theory of statistical mechanics, where they postulate in addition to PH and the dynamical equations (such as (\ref{HE}) and (\ref{H})), a probabilistic distribution of  the initial microstate of the universe: 
\begin{description}
  \item[Statistical Postulate (SP)] At the temporal boundary of the universe when PH applies, the probability distribution of the  microstate of the universe is given by the uniform one (according to the natural measure) that is supported on the macrostate of the universe (compatible with PH).
\end{description}

One may  also understand it in terms of typicality: that we regard the initial probability distribution to pick out a measure of almost all or the overwhelming majority---a measure of typicality \citep{goldstein2001boltzmann, goldstein2012typicality}. On this way of thinking, SP says the following: 
\begin{description}
  \item[SP'] At the temporal boundary of the universe when PH applies, the initial microstate of the universe is typical inside the macrostate of the universe (according to the natural measure of typicality).
\end{description}
On the basis of SP', one can then explore what the theory says about typical histories and apply it to our universe. 

A similar probabilistic boundary condition appears in Bohmian mechanics, where one can interpret the initial probability distribution of particle configuration as representing a typicality measure:
\begin{equation}\label{QEH}
\rho_{t_0} (q) = |\Psi(q, t_0)|^2
\end{equation}
where $t_0$ is when PH applies and $\Psi(q, t_0)$ is the wave function of the universe at $t_0$. Based on this measure, almost all worlds governed by Bohmian mechanics will exhibit the Born rule. (For more details, see \cite{durr1992quantum} and \cite{sep-qm-bohm}.)

In fact, it is also possible to interpret  stochastic dynamics as yielding a typicality measure: the GRW theory specifies a probability distribution over entire histories of the quantum states, and what matters is the behavior of ``almost all'' of those possible histories.

Probability measures and typicality measures are not straightforwardly understandable in terms of  MinP: it is not clear how they should be understood in terms of constraints. The difficulty is greater for stochastic dynamics. On the typicality approach, one has the option to regard the measures picked out by the probabilistic boundary conditions as referring to something methodological instead of nomological---how in practice one decides whether a law is supported or refuted by evidence. However, the probabilities in the stochastic dynamics are clearly nomological and not just a methodological principle of theory choice. 

  \cite[sect.3.3.3]{chenandgoldstein} discuss five interpretive strategies that are available on MinP. \cite{barrett2023algorithmic} have recently proposed a constraint account of probabilistic laws, where the nomic constraint is expressed in terms of Martin-L\"of randomness. The proposal is attractive from the perspective of MinP, since it provides a unified basis for understanding all candidate laws in terms of constraining laws.  However, there is much room for future work. As I shall discuss later, the problem of probabilistic laws is difficult on all accounts of laws.  Solving it may turn on questions about the relation between probability and typicality, and their relation to physical possibility.

\subsection{Summary}
MinP is an intelligible and attractive proposal for understanding fundamental laws of nature. It vindicates the non-Humean conviction that laws govern while remaining flexible enough to accommodate the variety of kinds of laws entertained in physics. In particular, it does not require that laws presume a fundamental direction of time.  MinP illuminates metaphysics but is not unduly constrained by it.

\section{Other Accounts}

In this section, I  survey five influential accounts of laws and compare them to MinP: Humean Reductionism, Platonic Reductionism, Aristotelian Reductionism, Langean Reductionism and Maudlinian Primitivism. With the exception of Humean Reductionism, they are more restrictive than MinP because of their metaphysical posits. Moreover, two of them--Aristotelian Reductionism and Maudlinian Primitivism--are explicitly committed to a fundamental direction of time. 

\subsection{Humean Reductionism}
We start with  Humean Reductionism, a metaphysically austere account that is as flexible as MinP (if not more so). On this view,  laws do not govern but merely summarize what actually happens in the world. Fundamental reality consists solely in the \textit{Humean mosaic}, a concrete example of which is a 4-dimensional spacetime occupied by particles and fields. At the fundamental level, laws of nature do not exist and do not move stuff around.  Laws are derivative of and ontologically dependent on the actual Humean mosaic.  The laws are the way they are \textit{because of} what the actual trajectories of particles and histories of fields are, not the other way around, in contrast to the governing picture of laws, such as the one given by MinP.  On MinP; the patterns in the Humean mosaic are ultimately unexplained by the laws; on Humean Reductionism, the laws are ultimately explained by the Humean mosaic, which in turn is not really explained by anything. 

Following Ramsey, \cite{LewisCounterfactuals, LewisNWTU, lewis1986philosophical} proposes a ``best-system'' analysis of laws that shows how laws can be recovered from the Humean mosaic. The basic idea is that laws are certain regularities of the Humean mosaic. However, not any regularity is a law, since some are accidental. One needs to be selective about which regularities to count as laws. Lewis suggests we pick those regularities in the best system of true sentences about the Humean mosaic. The strategy is to consider various systems (collections) of true sentences about the Humean mosaic and pick the system that strikes the best balance among various theoretical virtues, such as simplicity and informativeness. 

For a concrete example, let the Humean mosaic (the fundamental ontology) be a Minkowski spacetime occupied by massive, charged particles and an electromagnetic field. The locations and properties of those particles and the strengths and directions of the field at different points in spacetime is the matter distribution, which corresponds to the local matters of particular fact.  Suppose the matter distribution is a solution to Maxwell's equations. Consider three systems of true statements (characterized below using the axioms of the systems) about this mosaic:

\begin{itemize}
  \item System 1: \{Spacetime point $(x_1, y_1, z_1, t_1)$ has field strengths $E_1$ and $B_1$ with directions $\vec{v}_1$ and $\vec{v}_{1'}$   and is occupied by a particle of charge $q_1$;  spacetime point $(x_2, y_2, z_2, t_2)$ has field strengths $E_2$ and $B_2$ with directions $\vec{v}_2$ and $\vec{v}_{2'}$   and is not occupied by a charged particle; ......\}
  \item System 2: \{``Things exist.''\} 
  \item System 3: \{Maxwell's equations and the Lorentz force law\}
\end{itemize}
System 1 lists all the facts about spacetime points one by one. It has much informational content but it is complicated. System 2 is just one sentence that says there are things but does not tell us what they are and how they are distributed. It is extremely simple but has little informational content. 
System 3 lists just five equations of Maxwellian electrodynamics and the Lorentz force law. It has less information about the world than System 1 but has much more than System 2. It is more complicated than System 2 but much less so than System 1. 
System 1 and System 2 are two extremes; they have one virtue too much at the complete expense of the other. In contrast, System 3 strikes a good balance between simplicity and informativeness. System 3 is the best system of the mosaic. Therefore, according to the best-system analysis, Maxwell's equations and the Lorentz force law are the fundamental laws of this world. 

On this approach and unlike on MinP, laws are not fundamental facts that govern the universe, but are merely descriptive of and derivative from the Humean mosaic. Laws do not push or pull things, enforce behaviors, or produce the patterns. Laws are just winners of the competition among systematic summaries of the mosaic. \cite{BeebeeNGCLN} calls it the ``non-governing conception of laws of nature.'' Laws are merely those generalizations which figure in the most economical true axiomatization of all the particular matters of fact that happen to obtain.

Despite the simplicity and appeal of Lewis's analysis, there is an obstacle.  The theoretical virtue of simplicity is language-dependent. For example, suppose there is a predicate $F$ that applies to all and only things in the actual spacetime. Consider the following system: 
\begin{itemize}
  \item System 4: \{$\forall x F(x)$\}
\end{itemize}
This is informationally equivalent to System 1 and more informative than System 3, and yet it is simpler than System 3. If we allow competing systems to use predicate $F$, there will be a system (namely System 4) that is overall better than System 3. Given the best-system analysis, the actual laws of the mosaic would not be Maxwell's equations but ``$\forall x F(x)$.'' To rule out such degenerate systems, Lewis places a restriction on language. Suitable systems that enter into the competition can invoke predicates that refer to only natural properties. For example, the predicate ``having negative charge'' refers to a natural property, while the disjunctive predicate ``having negative charge or being the Eiffel Tower'' refers to a less natural property. Some properties are perfectly natural, such as those invoked in fundamental physics about mass, charge, spacetime location and so on. It is those perfectly natural properties that the axioms in the best system must refer to.  The predicate $F$ applies to all and only things in the actual world, which makes up an ``unnatural'' set of entities. $F$ is not perfectly natural; it is an example of a \textit{gruesome} predicate connected to Goodman's problem of induction (\S2.6). Hence, System 4 is not suitable. The requirement that the axioms of the best system refers only to perfectly natural properties is an important element of Lewis's Humeanism. 

Over the years, Lewis and his followers have, in various ways, extended and modified the best-system analysis of laws on Humean Reductionism. I have elsewhere called the updated view \textit{Reformed Humeanism about Laws}, but for simplicity here I will just call it the Best System Account (BSA):

\begin{description}
  \item[Best System Account (BSA)] The fundamental laws are the axioms of the best system that summarizes the mosaic and optimally balances simplicity, informativeness, fit, and degree of naturalness of the properties referred to. The mosaic contains only local matters of particular facts, and the mosaic is the complete collection of  fundamental facts. 
  \end{description}
BSA can accommodate various kinds of laws of nature. Without going into too much detail, we note the following features: 

   \textit{1. Chance}. Although chance is not an element of the Humean mosaic, it can appear in the best system. Humeans can introduce probability distributions as axioms of the best system \citep{LewisSGOC}. This works nicely for stochastic theories such as the GRW theory.  Humeans can evaluate the contribution of the probability distributions by using a new theoretical virtue called \textit{fit}. A system is more fit than another just in case it assigns a higher (comparative) probability than the other does the history of the universe. For certain mosaics, the inclusion of probability in the best system can greatly improve the informational content without sacrificing too much simplicity. Hence, fit can be seen as the probabilistic extension of informativeness.  Humeans can also allow what is called ``deterministic chance'' \citep{loewer2001determinism}.  Take a deterministic Newtonian theory of particle motion and add to it PH and SP, which can be represented as a uniform probabilistic distribution, conditionalized on a low-entropy macrostate of the universe at $t_0$. The Humean account of chance (both stochastic and deterministic) is arguably one of the simplest and clearest to date, but it still faces the Big Bad Bug \citep{LewisHSD} and the zero-fit problem \citep{elga2004infinitesimal}. 

  \textit{2. Flexibility with respect to the forms of laws}. Humean Reductionism is entirely flexible regarding the forms of laws. 
  Every example discussed in \S3.3 can be regarded as a best-system law that figures in the optimal summary of the Humean mosaic.  
   \cite{LewisNWTU} maintains that ``only the regularities of the system are to count as laws'' (p.367).  However, there is no reason to limit the Humean account to laws about general facts \citep{callender2004measures}. This flexibility is a significant advantage Humean Reductionism has over some other accounts of laws \citep{loewer2012two}.

   \textit{3. Flexibility with respect to perfect naturalness}. For Lewis, perfect naturalness is a property of properties. Perfectly natural properties pick out the same set of things as Armstrong's theory of sparse universals (more on that in \S2.2). However, the chief motivation for Lewis's use of perfect naturalness is to rule out systems that use ``gruesome'' predicates. If that is the issue, as \cite{hicksschaffer} suggest, we can simply require that ``degree of naturalness'' of the predicates involved be a factor in the overall ranking of competing systems, and the best system should also optimally balance degree of naturalness of the predicates together with the rest of the theoretical virtues, such as simplicity, informativeness, and fit.  The flexibility with respect to perfect naturalness also allows the best system to refer to non-fundamental properties such as entropy, as may be necessary if the Past Hypothesis is a fundamental law. 

   

\textit{Comparisons with MinP.}  Humean Reductionism and MinP are similar in several respects. First, neither requires a fundamental direction of time, and both permit a reductionist understanding of it. Second, both views are flexible enough to accommodate the distinct kinds of laws entertained in physics. Third, both views emphasize the importance of simplicity (and other super-empirical virtues) in laws and scientific explanations.  

Let me turn to their differences. First, on Humean Reductionism, the patterns in the Humean mosaic have no ultimate explanation; after all, the mosaic grounds what the laws are. Many reject Humean Reductionism for that reason.\footnote{There is an on-going debate about the status and nature of explanation on Humeanism. For some examples, see \cite{loewer2012two, lange2013grounding} and \cite{emery2019laws, emery2023naturalism}. } On MinP, suitable explanations of the patterns must not be merely summaries of the mosaic. On MinP, fundamental laws are metaphysically fundamental facts that exist in addition to the mosaic.  They govern the mosaic and explain its patterns by constraining it in an illuminating manner.  

Second, they also differ regarding the modal profile of laws. This has been much discussed in the literature (see for example \cite{carroll1994laws} and \cite{MaudlinMWP}). On MinP, since fundamental laws are primitive facts, there can be a physically possible world corresponding to an empty Minkowski spacetime governed by the Einstein equation. However, on Humean Reductionism, that world is one where the simplest summary is just the laws of special relativity, and it is impossible to have such a world where the law is the Einstein equation \cite[pp. 67-68]{MaudlinMWP}. To allow two worlds with the same mosaic (empty Minkowski spacetime) but different laws (laws of special relativity and those of general relativity), \x{which is accepted in scientific practice, is to endorse non-supervenience of the laws on the mosaic. Therefore, Humean Reductionism seems to be in conflict with scientific practice while MinP is not.}\footnote{\x{See \cite{roberts2008law} for a Humean account of laws based on a contextualist semantics that may alleviate this worry.}}  

Third, some Humeans suggest that,  precisely because of the metaphysical difference concerning supervenience, Humeanism has an \textit{epistemological advantage} over non-Humean views such as MinP \citep{earman2005contact}. Since Humean laws are supervenient on the mosaic, and since the mosaic is all that we can empirically access (we do not directly see the laws), we are in a better position to determine laws on Humeanism than on non-Humeanism. This argument becomes less convincing once we remind ourselves that \textit{we are not given the mosaic}; the mosaic in modern physics is as theoretical as the laws. I return to this point in \S5. 

Finally, Humean Reductionism faces what \cite{LewisHSD} calls the problem of ``ratbag idealism.'' Since the best systematization is constitutive of lawhood, and if what counts as best depends on us, lawhood may become mind-dependent. In contrast, on MinP, fundamental laws are what they are irrespective of our psychology and judgments of simplicity and informativeness. Even though the epistemic guides provide some guidance  for discovering and evaluating them, they do not guarantee arrival at the true fundamental laws. Moreover, changing our psychology or judgments will not change which facts are fundamental laws. Hence, MinP respects our conviction about the objectivity and mind-independence of fundamental laws. For more discussions, see \S7. 

\subsection{Platonic Reductionism}

With Humean Reductionism,  nothing ultimately explains the patterns in the Humean mosaic. 
Those with a governing conception of laws may seek to find a deeper explanation. In virtue of what is every massive particle in the world behaving according to the formula $F=ma$? What, if anything, enforces the pattern and makes sure nothing deviates from it? 
On MinP, it is the laws themselves. 

\cite{DretskeLN}, \cite{tooley1977nature}, and \cite{ArmstrongWIALON} propose a different non-Humean account of governing laws, one that seeks to understand them in terms of \textit{universals}. The universals that they accept  are  in addition to things in the Humean mosaic. They are ``over and above'' the Humean mosaic. In traditional metaphysics, universals are repeatable entities that explain  the genuine similarity of objects. Let us start with some mundane examples. Two cups are genuinely similar in virtue of their sharing a universal \textit{Being a Cup}. The universal is something they both instantiate and something that explains their genuine similarity. A cup is different from a horse because the latter instantiates a different universal \textit{Being a Horse}. Now, those universals are not fundamental, and they may be built from more fundamental universals about physical properties.  Dretske, Tooley, and Armstrong use universals to provide explanations in science. For them, the paradigm examples are universals that correspond to fundamental physical properties, such as mass and charge.  On their view, laws of nature hold because of a certain relation obtaining among such universals. This theory of laws has  connection to Plato's theory of forms.\footnote{For an overview of Plato's theory of forms, see \cite{sep-plato}.} We thus call it \textit{Platonic Reductionism}.\footnote{In the literature it is sometimes called the DTA account of laws or the Universalist account of laws. Calling it Platonic \textit{reductionism} may be controversial. But see the discussion in \cite[appendix A1]{carroll1994laws}.   }
 
Consider the world where $F=ma$ holds for every massive particle. In such a world, any particle with mass $m$ instantiates the universal \textit{having mass $m$}, any particle under total force $F$ instantiates the universal \textit{being under total force $F$}, and any particle with acceleration $F/m$ instantiates  the universal \textit{having acceleration $F/m$}. The universals are multiply instantiated and repeated, as there are many particles that share the same universals. Those universals give unity to the particles that instantiate them.  The theory also postulates, as a fundamental fact, that the universal \textit{having mass $m$} and the universal \textit{being under total force $F$} necessitate the universal \textit{having acceleration $F/m$}. Hence, if any particle instantiates  \textit{having mass $m$} and \textit{being under total force $F$}, then it has to instantiate \textit{having acceleration $F/m$}. It follows that every particle has to obey $F=ma$.\footnote{We note that this example about $F=ma$ does not exactly fit in Armstrong's schema of ``All F's are G.'' See \cite[ch.7]{ArmstrongWIALON} for a proposal for accommodating  ``functional laws.'' } This adds the necessity and the oomph that are missing in Humean Reductionism. 

With Platonic Reductionism, the regularity is  explained by the metaphysical postulate of universals and the necessitation relation $N$ that hold among universals.  Following \cite{hildebrand2013can}, we may summarize it as follows: 

\begin{description}
  \item[Necessitation] For all universals $F$ and $G$, $N(F,G)$ necessitates the regularity that all $F$s are $G$s. 
\end{description}
Some clarificatory remarks:

   \textit{1. Universals}. (i) The appeal to universals is indispensable in this theory of laws. The theory is committed to a fundamental ontology of objects (particulars) and a fundamental ontology of universals.  Hence, Platonic Reductionism  is  incompatible with nominalism about universals.  
   (ii) Defenders such as Armstrong appeal to a sparse theory of universals, where the fundamental universals correspond to the fundamental properties we find in fundamental physics. The sparse universals correspond to the perfectly natural properties that Lewis invokes in his account. Consider Lewis's example of the predicate $F$ that corresponds to the property of all and only things in the actual world. For Armstrong,``$\forall x F(x)$'' does not express a fundamental law because objects with property $F$ are not genuinely similar, and $F$ is a property that does not correspond to one of the fundamental, sparse universals. 

   \textit{2. Necessity}. (i) The necessity relation among universals is put into the theory by hand. It is a postulate that such a relation holds among universals and does necessitate regularities. (It is also postulated that the relation among universals is itself a universal.) To some commentators, it is unclear why the postulate is justified \cite[p.366]{LewisNWTU}.
In response, a defender of Platonic Reductionism may take the necessity relation simply as a primitive and stipulate its connections to regularities \citep{schaffer2016business}.  
 

 (ii) \cite[p.172]{ArmstrongWIALON} understands probabilistic laws as giving ``a probability of necessitation'' between two universals. 
 What is ``a probability of a necessitation?'' Conceptually, whether $F$ necessitates $G$ seems like a matter that does not admit of degree. What does this probability mean, and how does it relate to actual frequencies and why should it constrain our credences? Even if one accepts the intelligibility of the necessitation relation, one may be unwilling to accept the intelligibility of objective probability of a necessitation and one may be puzzled by how the probability of a necessitation can explain the regularities. 
 

On Platonic Reductionism, it is unclear how we should think about the direction of time.  Even though there is a strong connection between the necessitation relation $N$ and causation, it does not seem that the main defenders build the direction of time into $N$. 
Nevertheless, if Platonic Reductionism does not have room for  treating the Past Hypothesis as a fundamental law, it may need to invoke a fundamental direction of time for worlds like ours. Perhaps Platonic Reductionism is best paired with a primitivism about the direction of time.  

\textit{Comparisons with MinP.}  Platonic Reductionism and MinP agree  that there are governing laws that do not supervene on the Humean mosaic, but disagree on whether governing laws should be analyzed in terms of or are reducible to relations among universals. 

While Platonic Reductionism is ontologically committed to fundamental universals, MinP is not. I do not think that universals offer additional explanatory benefits. The motivating idea of Platonic Reductionism is that universals are  properties that genuinely similar objects share, and it is partly in virtue of the universals shared by those objects that the objects behave in the same way everywhere and everywhen. The metaphysics of $N$ is a complicated business, and it seems to create more mystery than it dispels. In contrast, on MinP  there are fundamental laws that govern the world by constraining the physical possibilities.  Explanation in terms of simple laws seems clear enough to vindicate the non-Humean intuition that there is something more than the mosaic that governs it. Moreover, MinP is compatible with various metaphysical views about properties such as realism and nominalism. A realist attitude towards laws does not require a realist attitudes towards properties. 

Unlike MinP, Platonic Reductionism places restrictions on the form of fundamental laws. On Platonic Reductionism, all laws need to be recast in the form of relations among universals, and it is unclear how to do so for laws in modern physics. (See also \cite[sect 3.2]{hildebrand2021nomological}.)  
Consider a differential equation that expresses a candidate fundamental law such as (\ref{HE}).  What are the universals that they actually relate? Assuming that velocity and acceleration are derived quantities,  what are the universals that correspond to the derivatives on either side of the equations? Armstrong argues that universals must be instantiated in some concrete particulars. As Wilson observes (p.439), differential equations conflict with Armstrong's principle about the instantiation of universals, as the values of the derivatives are calculated from values possessed by non-actual states (those in the small neighborhood around the actual one) that are not  instantiated. In contrast, MinP has no difficulty accommodating laws expressed by differential equations. 

Moreover, some candidate fundamental laws involve properties that do not seem to correspond to universals. For example, PH  applies to only one moment in time. As such, it is a spatiotemporally restricted law that seems in tension with the approach involving universals (universal, repeatable, and multiply instantiated).  \cite{tooley1977nature} considers an example of Smith's Garden, and there he seems open to accept spatiotemporally restricted laws if they are significant enough. One can devise semantic tricks to understand them in terms of universals,  but it is hard to see what the point is. In contrast, MinP has no difficulty accommodating spatiotemporally restricted laws; they can function perfectly as constraints on the universe that are about specific places or times.

\subsection{Aristotelian Reductionism}

The next view is most commonly associated with contemporary defenders of dispositional essentialism.  On this view, laws are not fundamental entities; neither do they govern the world in any metaphysically robust sense. Laws do not push or pull things around. Instead, the patterns we see are explained by the fundamental properties that objects instantiate. Those properties are the seats of metaphysical powers, necessity, and oomph. Those properties make objects, in a certain sense, ``active'' \citep[p.1]{ellis2001scientific}. Such properties are often called  ``dispositions,'' and also sometimes called  ``powers,'' ``capacities,'' ``potentialities,'' and ``potencies.''\footnote{For an overview of the metaphysics of dispositions, see \cite{sep-dispositions}.} However, they are different from the universals in Platonic Reductionism or the natural properties in Humean Reductionism, which may be viewed as ``passive.''    If there are any laws (and there is an internal debate about this question among defenders of this fundamental dispositional ontology), they derive from or originate in the fundamental dispositions of material objects. 

Roughly speaking, objects with dispositions have characteristic behaviors (also called manifestation) in response to certain stimuli \citep[p.3]{BirdNM}.  For example, a glass has a disposition to shatter when struck; an ice cube has a disposition to melt when heated; salt has a disposition to dissolve when put into water. On this view, fundamental properties are similarly dispositional: negatively charged particles have a disposition to attract positively charged particles; massive particles have a disposition to accelerate in a way that is proportional to the total forces on them and inversely proportional to their masses.  Moreover, a dispositional essentialist holds that some properties have dispositional essences, i.e. their essences can be characterized in dispositional terms.\footnote{Some, such as \cite{BirdNM}, go further and claim that all perfectly natural properties in Lewis (1986)'s sense or all sparse universals in Armstrong (1983)'s sense have dispositional essences.}

Among those who endorse a dispositionalist fundamental ontology, not everyone accepts that fundamental laws, which are usually taken to be universally valid and always true, arise from dispositions.   For example, \cite{cartwright1983, cartwright1994nature}  and \cite{mumford2004laws} deny the need for such laws.  Nevertheless, the dispositional essentialists need not abandon laws. They can maintain that laws supervene on or reduce to dispositions. Because of its Aristotelian roots \citep{ellis2014philosophy}, we call such a view \textit{Aristotelian Reductionism} about laws.\footnote{Many defenders of this view suggest that even though it has roots in Aristotle, it is not committed to many aspects of Aristotelianism.} 
Aristotelian Reductionists maintain that (i) the metaphysical powers, necessity, and oomph reside in the fundamental dispositions; (ii) laws are metaphysically derivative of the dispositions; (iii) laws are metaphysically necessary. 




How are  laws derived from dispositions?  Bird proposes that we can derive laws from certain counterfactual conditionals associated with dispositional essences. A more recent approach is that of \cite{demarest2017powerful, demarestMC} and \cite{kimpton2017humean} that seek to combine a dispositional fundamental ontology with a best-system analysis of lawhood. Here we focus on the approach of Demarest. She proposes that dispositions (she follows Bird and calls them potencies) do metaphysical work. They produce their characteristic behaviors, resulting in patterns in nature.  Their characteristic behaviors, in different possible worlds, can be summarized in simple and informative axiomatic systems, and the best one contains the true laws of nature. That is like Humean Reductionism except that (i) Demarest's fundamental ontology includes potencies and (ii) the summary is  not of just the actual distribution of potencies but also all merely possible ones. In this way, her proposal may be an elaboration of Bird's suggestion that we can derive laws from potencies, though she does not rely on counterfactuals. 


In contrast to Humean Reductionism, here the patterns are ultimately explained by the potencies. How do potencies explain? Demarest appeals to dynamic production \citep[pp.51-52]{demarest2017powerful}. 
The potencies at an earlier time explain how things move at a later time by dynamically producing, determining, or generating the patterns.  Demarest's view seems committed to a fundamental direction of time. The account of dynamic explanation presupposes a fundamental distinction between past and future, i.e. between the initial and the subsequent states of the world. The initial arrangement of particles and potencies  metaphysically ground subsequent behaviors of particles.  The commitment of a fundamental direction of time does not seem optional on her view. 

Moreover, the metaphysical framework of fundamental dispositions already seems committed to a fundamental direction of time, independently of the issue of laws. For example, it is natural to interpret the discussions by Ellis, Bird, Mumford as  suggesting that the manifestation of a disposition cannot be temporally prior to its stimulus, which presupposes a fundamental direction of time.\footnote{In contrast, \cite{vetter2015potentiality} is open to a temporally symmetric metaphysics but assumes temporal asymmetry in her account of dispositions (which she calls potentialities).} Therefore, although Aristotelian Reductionism does away with the governing conception of laws, the view seems committed to a fundamental direction of time twice over. 

\textit{Comparisons with MinP.} Aristotelian Reductionists do not think that laws govern in a metaphysically robust sense. In contrast, MinP vindicates the conviction that laws do.  Aristotelian Reductionism is committed to a fundamental ontology of dispositions. MinP is not. Most physicists today may be unfamiliar with the concept of fundamental dispositions. They are familiar with the concept of fundamental laws and how they figure in various scientific explanations. Hence, MinP seems more science-friendly.  

Moreover, it is natural to read dispositional essentialists such as Bird, Mumford, and Ellis as having an implicit commitment to a fundamental direction of time. Demarest's account is more explicit in linking the dispositional essentialist ontology and the account of nomic explanations to that of dynamic metaphysical dependence, or what we call dynamic production. As noted in \S3.3, it is difficult to understand how dynamic production works even in simple cases such as Hamilton's equations and much less in relativistic spacetimes. Requiring dynamic production presumably rules out theories that permit closed timelike curves, as well as purely spatial laws, or even worlds for which spacetime is emergent.  In contrast, MinP is not committed to a fundamental direction of time, and MinP is entirely open to those possibilities (even though we may have other considerations, beyond the conception of laws, to not consider them). 

Finally, there are problems specific to accounts (such as Bird's) that analyze laws in terms of dispositions. \cite{BirdNM} lists four problems (p.211): (i) fundamental constants, (ii) conservation laws and symmetry principles,  (iii) principles of least action, and (iv) multiple laws relating distinct properties. Problem (i) arises because slight differences in the constants do not require the properties to be different; problem (ii)  because conservation laws and symmetry principles do not seem to be manifestations of dispositions; problem (iii)  because the principles seem to commit to the physical possibilities of alternate histories, something not allowed on dispositional essentialism; problem (iv) because a third law relating two properties will not be the outcome of the dispositional natures of those properties. Such problems do not  arise on MinP. 

\subsection{Langean Reductionism}

Lange (\citeyear{lange2009laws}, \citeyear{lange2005laws}, \citeyear{lange1999laws}) develops a non-Humean account where laws are explained in terms of counterfactuals.  He suggests that counterfactuals (instead of laws as on MinP) should be regarded as ontologically basic.  Let us call the view \textit{Langean Reductionism}. It is also more restrictive than MinP. 

 To unpack his views, we need to understand his definition of lawhood. Let  us denote the set of (first-order) laws of nature together with their logical consequences as $\Lambda$. It is understood to include only what Lange (2009, p.17) calls ``sub-nomic truths,'' those that do not directly make claims about lawhood, such as $F=ma$, but not \textit{it is a law that} $F=ma$. On Lange's account (p.42),  $\Lambda$ stands in a special relationship to counterfactuals. He proposes that  $\Lambda$ is the ``largest nonmaximal sub-nomically stable set'' of truths about the universe.  Roughly speaking, $\Lambda$ is the largest set that is (1) not the set of all (sub-nomic) truths about the universe, i.e. nonmaximal, and (2) counterfactually stable under any (sub-nomic) supposition that is consistent with $\Lambda$, i.e. sub-nomically stable. To satisfy (2), $\Lambda$ must be such that, for any sub-nomic proposition $p$ that does not make claims about lawhood or conflict with $\Lambda$, every member of $\Lambda$ would still be true if $p$ were true.   \footnote{For a more precise and complete definition of (2), which involves nested counterfactuals, see  \cite[p.29]{lange2009laws}.} 
 
 This special relationship between $\Lambda$ and counterfactuals, on Lange's view, provides a principled and sharp distinction between laws and accidents. It allows us to define laws from counterfactuals in a non-circular way, achieving one of the central aims of Lange's project and distinguishing it from all other accounts in the literature. 

The nonmaximality condition plays a crucial role here.  Lange shows that every set containing accidents, with the possible exception of the set of all sub-nomic truths, lacks the special property of sub-nomic stability.  As Lange explains (p.30), on many logics of counterfactuals, the set of all sub-nomic truths (including all accidents) also forms a sub-nomically stable set, albeit a \textit{maximal} one (because it cannot be expanded without becoming inconsistent). Without the further constraint that laws cannot form a maximal set, there will be no principled and sharp distinction between laws and accidents, for sub-nomic stability does not guarantee lawhood.  One can define laws to be the stable set that contains no accidents, but that would be circular, a feature he wants to avoid. What does the trick is defining laws to be the largest \textit{nonmaximal} sub-nomically stable set.  

What make a proposition a law of nature, i.e. the lawmakers, are the counterfactuals, which are metaphysically fundamental.  This reverses the direction of metaphysical explanation, as we usually think that laws support counterfactuals and not the other way around (\S2.3). However, Lange does not deny that sometimes our knowledge of laws can be used to evaluate counterfactuals, even though laws depend on counterfactuals.  

This view is arguably compatible with the Past Hypothesis being a fundamental law. Hence, it is compatible with a reductionist understanding of the direction of time, and it does not assume that laws must dynamically produce the states of the universe in order to govern. 


\textit{Comparisons with MinP.} Langean reductionism is one of the more flexible non-Humean accounts of laws, as it can accommodate a wide variety of laws.  However, it is still more restrictive than MinP.  Because Lange requires that $\Lambda$ be nonmaximal,  Langean Reductionism is incompatible with even the metaphysical possibility of strong determinism. 

To see the conflict, let us suppose strong determinism is metaphysically possible. Consider a possible world $w$ where strong determinism is true. The laws of $w$ is compatible with only $w$. Thus, the $\Lambda$ at $w$, being logically closed, entails all (sub-nomic) truths at $w$. That makes $\Lambda$ the \textit{maximal} sub-nomically stable set at $w$,  contradicting the requirement that $\Lambda$ be nonmaximal. Hence, defining  $\Lambda$ to be nonmaximal rules out the \textit{metaphysical possibility} of strong determinism. 


I regard that as a significant and under-appreciated cost of Lange's account.  Strong determinism is logically consistent and conceptually coherent.  On what grounds are we entitled to dismiss it as metaphysically impossible? Moreover, there can be scientifically motivated, empirically adequate, and theoretically virtuous candidates for  laws that are strongly deterministic and provide new insights into the foundations of physics (\S2.4; \S3.3). Ruling out strong determinism by fiat seems contrary to the methodology of naturalistic metaphysics. I think the lesson here is that we should rethink the original motivation for a non-circular distinction between laws and accidents and modify Lange's account in light of the possibility of strong determinism.

In contrast, strong determinism is not only allowed on MinP but  serves as a good example of how laws constrain the universe---strongly deterministic laws rules out every world except the actual world. If such laws are simple, they provide strong and compelling explanation for every event in spacetime. 

\subsection{Maudlinian Primitivism}
All previous accounts surveyed in this section  attempt to reduce laws to something else, such as the Humean mosaic, universals, dispositions, and counterfactuals. An alternative view is simply to take laws as ontological primitives. An influential primitivist account is developed by  \cite{MaudlinMWP}. His version of primitivism, which in one aspect is similar to MinP, comes with more metaphysical commitments than MinP. 

 As a primitivist about laws, he suggests that we should not analyze or reduce laws into anything else. Laws are metaphysically fundamental; they are primitive entities that do not supervene on other entities. To have a sufficiently explanatory metaphysical theory, our fundamental ontology needs to include not only spatiotemporal objects but also laws that govern them. Maudlin rejects any reduction or deeper analysis of laws. 

Maudlin is also committed to primitivism about the direction of time: that the distinction between past and future is metaphysically fundamental and not reducible to anything else. There is in effect a fundamental arrow or orientation at every spacetime point that points to the future.   Maudlin combines the two commitments into a metaphysical package (p.182).  

The two posits are closely connected on his account. For Maudlin, laws produce or generate later states of the world from earlier ones. In this way, via the productive power of the laws, subsequent states of the world (and its parts) are explained by earlier ones and ultimately by the initial state of the universe. It is this productive explanation that is central to his account.  Dynamic production is closely related to causation, and just like (paradigm cases of) causation it is time asymmetric. Future states are dynamically produced from earlier states but not vice versa.  This, for example, allows Maudlin's account to vindicate a widespread intuition about Bromberger's flagpole. The shadow is produced by the circumstances and the length of the pole (together with the laws). Although we can deduce from the laws the pole length based on the circumstances and the shadow length, the pole length is not produced by them. Hence, given the laws, the pole length and the circumstances explain, but are not explained by, the shadow length. 
 
Maudlin suggests that his package yields an attractive picture by being closer to our initial conception of the world. 
He is committed to all three theses discussed in \S2.4: Only FLOTEs,  Dynamic Production, and Temporal Direction Primitivism. We may summarize Maudlin's metaphysical package as follows:
\begin{description}
  \item[Maudlinian  Primitivism] Fundamental laws are certain ontological primitives in the world.  Only dynamical laws (in particular, laws of temporal evolution) can be fundamental laws. They operate on the universe by producing later states of the universe from earlier ones, in accord with the fundamental direction of time. 
\end{description}
Maudlin allows there to be primitive stochastic dynamical laws---those laws that involve objective probability such as the GRW collapse laws. Hence, dynamic production need not be deterministic. An initial state can be compatible with multiple later states, determining only an objective probability distribution over those states. Perhaps the objective probability can be understood as \textit{propensity}, with stochastic production implying variable propensities of producing various states, in proportion to their objective probabilities and in accord with the direction of time. However, even if deterministic production is an intelligible notion, it is not clear that stochastic production or propensity is as intelligible. (Recall the earlier point about ``probabilistic necessitation'' in Platonic Reductionism.) 

\textit{Comparisons with MinP.} Maudlinian Primitivism and MinP agree that fundamental laws are metaphysically fundamental and that they govern. They disagree about how they do it. 

For Maudlin, dynamic production is essential, and every fundamental law needs to have the form of a dynamical law (in the narrow sense of a FLOTE) that can be interpreted as evolving later states of the universe from earlier ones. For laws to produce, they operate according to the fundamental direction of time, providing an intuitive picture close to our pre-theoretic conception of the world: ``the universe is generated from a beginning and guided towards its future by physical law'' (p.182). MinP is not committed to a fundamental direction of time; nor is it committed to dynamic production as how laws govern or explain. On MinP, explanation by simple constraint is good enough. Many candidate fundamental laws such as the Einstein equation are not (in and of themselves) FLOTEs that produce later states of the universe from earlier ones. For the same reason, PH, WCH, IPH, and NBWF cannot be Maudlinian laws. And neither can a purely spatial constraint such as Gauss's law\footnote{\x{In personal communication, Maudlin suggests that he now regards (\ref{Gauss}) as expressing a metaphysical analysis or a definition of $\rho$ in terms of the divergence of $\textbf{E}$. } } or the simple rule responsible for the Mandelbrot world. On MinP, they can all be understood as fundamental laws that express simple constraints. 

The difficulty with dynamic production is not just that it precludes certain candidate fundamental laws. It is also difficult to understand the notion itself. What does dynamic production mean and what are its relata? Does it relate instantaneous states or sets of instantaneous states of the universe? If it relates instantaneous states, it is hard to understand dynamic production even in paradigm examples of FLOTEs such as the one expressed by Hamilton's equations. (The initial data is not confined to a single moment in time, if we understand momentum as partly reducible to variations in positions over some time interval.)  The notion becomes even less natural in relativistic settings. 
Moreover, on a simple understanding of dynamic production, the beginning of the universe does  metaphysical work; it is what gets the entire productive enterprise started. However, for spacetimes with no temporal boundaries,   it is unclear where to start the productive explanation.  
In contrast, constraints operate on the entire spacetime, regardless of whether there is   an ``initial'' moment. Thus, MinP does not require a first moment in time.  (Perhaps a more sophisticated understanding of dynamic production does not either.)

\x{On MinP, even if the universe lacks a fundamental direction of time, we can still recover a notion of productive explanation at a non-fundamental level. For example, we can use PH to define a (non-fundamental) direction of time in the usual way: \textit{earlier} is defined as being closer to the time of PH, while \textit{later} is defined as being further away from that. We may regard FLOTEs as evolving earlier states of the universe into later ones (with respect to PH).  In such a universe, dynamic production may be metaphysically derivative. Still, we can contemplate a (non-fundamental) productive explanation of Bromberger's flagpole and vindicate the intuition that the pole length and the circumstances explain, but are not explained by, the shadow length.  Therefore, the intuitive picture behind Maudlinian Primitivism can be preserved even if there is no fundamental notion of dynamic production or a fundamental direction of time. }

\subsection{Summary}

We have looked at five accounts of laws that have been widely discussed in the literature, and compared them with MinP. All four non-Humean accounts surveyed here are more restrictive than MinP. Even though Humean Reductionism disagrees with MinP regarding the metaphysical status of laws, they agree on many first-order judgments about which equations may express laws. This suggests that the fundamental disagreement between Humeanism and non-Humeanism is much more subtle than is often recognized. 

\section{Simplicity}

Physical laws are strikingly simple, although there is no \textit{a priori} reason they must be so. I propose that nomic realists of all types (Humeans and non-Humeans) should accept that simplicity is a fundamental epistemic guide for discovering and evaluating candidate  laws. As it turns out, this epistemological principle  is independent of the metaphysical posits about laws in both Humeanism and non-Humeanism. 

\subsection{Nomic Realism and the Epistemic Gap}

Many physicists and philosophers are realists about physical laws. Call realism about physical laws \textit{nomic realism}. It contains two parts:
\begin{description}
  \item[Metaphysical Realism:] Physical laws are objective and mind-independent; more precisely, which propositions express physical laws are objective and mind-independent facts in the world.\footnote{A weaker version of metaphysical realism maintains that laws are not entirely mind-dependent. That would accommodate more pragmatic versions of the Humean best-system accounts including those of \cite{hicks2018dynamic, dorst2019towards, jaag2020making}, and \cite{callender2021humean}. The arguments below, with suitable adaptations, still apply.}
\end{description}

\begin{description}
  \item[Epistemic Realism:] We have epistemic access to physical laws; more precisely, we can be epistemically justified in believing which propositions express the physical laws, given the evidence that we will in fact obtain.\footnote{The terminology is due to \cite{earman2005contact}. Here I've added the clause ``given the evidence that we will in fact obtain.''  My version of epistemic realism is logically stronger than theirs. }  
\end{description}

Nomic realism gives rise to an apparent epistemic gap: if  laws are really objective and mind-independent, it may be puzzling how we can have epistemic access to them, since laws are not consequences of our observations. The epistemic gap can be seen as an instance of a more general one regarding theoretical statements on scientific realism \citep{sep-scientific-realism}. 

The accounts surveyed in \S4-5 all aspire to satisfy nomic realism. Let us focus on BSA and MinP. Do they vindicate epistemic realism? Their metaphysical posits, by themselves, do not guarantee epistemic realism. This should be clear on MinP. Since there is no metaphysical restriction on the form of the fundamental laws,  if they are entirely mind-independent primitive facts about the world, how do we know which propositions are the laws? However,  an analogous problem exists on BSA.  This claim may surprise some philosophers, as it is often thought that BSA has an epistemic advantage over non-Humean accounts like MinP, precisely because BSA brings laws closer to us. BSA defines laws in terms of the mosaic, and the mosaic is all we can empirically access \citep{earman2005contact}. 

The problem is that \textit{we are not given the mosaic}. Just like  laws, the mosaic entertained in modern physics is a theoretical entity that is not entailed by our observations. Our beliefs about its precise nature, such as the global structure of spacetime, its microscopic constituents, and the exact matter distribution, are as theoretical and inferential as our beliefs about  laws. They are all parts of a theory about the physical world. Just as MinP requires an extra epistemic principle to infer what the laws are,  BSA requires a similar principle to infer  what the mosaic is like. The latter, on BSA, turns out to be equivalent to a strong epistemic principle concerning what we should expect about the best system given \textit{our limited evidence}, which because of its limitation pins down neither the mosaic nor the best system. 

After all, on BSA laws are not summaries of our observations only, but of the entire spacetime mosaic constituted by the totality of microphysical facts, a small minority of which show up in our observations. The ultimate judge of which system of propositions is the optimal true summary depends on the entire mosaic, a theoretical entity. (For this reason, BSA should not be confused as a version of strict empiricism.) And in current physics, our best guide to the mosaic is our best guess about the  laws.  At the end of the day, MinP and BSA turn out to require the same super-empirical epistemic principle concerning  laws.  On neither account does the epistemic principle follow from the metaphysical posits about what laws are. 

To sharpen the discussion, let us suppose, granting Lewis's assumption of the kindness of nature \cite[p.479]{LewisHSD},  that given the mosaic $\xi$ there is a unique best system whose axioms express the fundamental law $L$: 
\begin{equation}
	L = BS(\xi)
\end{equation}
with $BS(\cdot)$ the function that maps a mosaic to its best-system law.\footnote{We might understand pragmatic Humeanism as recommending that we use another best-system function $BS'(\cdot)$ that is ``best for us.'' } Let us stipulate that for both BSA and MinP,  physical reality is described by a pair $(L, \xi)$. For both, we must have that $\xi \in \Omega^L$, with $\Omega^L$ the set of mosaics compatible with $L$. This means that $L$ is true in $\xi$.  On BSA, we also have that 	$L = BS(\xi)$. So in a sense, all we need in BSA is $\xi$; $L$ is not ontologically extra. But it does not follow that BSA and MinP are relevantly different when it comes to epistemic realism. 

Let $E$ denote our empirical evidence consisting of our observational data about physical reality. Let us be generous and allow $E$ to include not just our current data but also all past and future data about the universe that we in fact gather.  There are two salient features of $E$: 
\begin{itemize}
  \item $E$ does not pin down a unique $\xi$. There are different candidates of $\xi$ that yield the same $E$. (After all, $E$ is a spatiotemporally partial and macroscopically coarse-grained description of $\xi$.) 
  \item  $E$ does not pin down a unique $L$.  There are different candidates of $L$ that yield the same $E$. (On BSA, this is an instance of the previous point; on MinP, this is easier to see since $L$ can vary independently of $\xi$, up to a point.)
\end{itemize}
Hence, on BSA, just as on MinP, $E$ does not pin down $(L, \xi)$. There is a gap between what our evidence entails and what the laws are. Ultimately, the gap can be bridged by adopting  simplicity (among other super-empirical virtues) as an epistemic guide. Nevertheless, it helps to see how big the gap is so that we can appreciate how much work needs to be done by simplicity and other epistemic guides.\footnote{It is worth contrasting the current setup with the influential framework suggested by \cite{hall2009humean, hall2015humean}.  In one way of fleshing out the core idea of BSA, Hall imagines a Limited Oracular Perfect Physicist (LOPP) who has as her evidence all of $\xi$ and nothing else. Her evidence $E_{\text{LOPP}}$ contains vastly more information than $E$.  On BSA, $E_{\text{LOPP}}$  pins down $(L, \xi)$. Actually, this is still incorrect. On BSA, we also need the assumption that  $E_{\text{LOPP}}$ corresponds to the entire spacetime, and the mosaic does not ``continue'' after $E_{\text{LOPP}}$. This subtlety has not been sufficiently appreciated in the literature.  Notice that $E_{\text{LOPP}}$ is as theoretical for Humeans as for non-Humeans. The Humean's best guess about what is in $E_{\text{LOPP}}$ depends on her expectation about what $L$ looks like given $E$.  }

The epistemic gaps can be illustrated by considering cases of empirical equivalence.  If different laws yield the same evidence, it is puzzling how we can be epistemically justified in choosing one over its empirically equivalent rivals, unless we rule them out by positing substantive assumptions that go beyond the metaphysical posits of nomic realism. Here I briefly mention three kinds of algorithms for generating empirical equivalents. For a more in-depth discussion, see \cite{chen2023simplicity}. 
\\

\noindent
\textbf{Algorithm A: Moving parts of ontology (what there is in the mosaic) into the nomology (the package of laws)}. 

   \textit{General strategy.} This strategy works on both BSA and MinP. Given a theory of physical reality $T_1=(L, \xi)$, if $\xi$ can be decomposed into two parts $\xi_1 \& \xi_2$,  we can construct an empirically equivalent rival $T_2 = (L\&\xi_1, \xi_2)$, where $\xi_1$ is moved from  ontology to nomology.   
 
\textit{Example.}  Consider the standard theory of Maxwellian electrodynamics, $T_{M1}$: 
  \begin{itemize}
  \item   Nomology: Maxwell's equations and Lorentz force law.
  \item Ontology: a Minkowski spacetime occupied by charged particles with trajectories $Q(t)$ and an electromagnetic field $F(x,t)$. 
\end{itemize}

Here is an empirically equivalent rival, $T_{M2}$: 
  \begin{itemize}
  \item   Nomology: Maxwell's equations, Lorentz force law, and an enormously complicated law specifying the exact functional form of $F(x,t)$ that appears in the dynamical equations.
  \item Ontology: a Minkowski spacetime occupied by charged particles with trajectories $Q(t)$.
\end{itemize}
Our evidence $E$ is compatible with both $T_{M1}$ and $T_{M2}$. The outcome of every experiment in the actual world will be consistent with $T_{M2}$, as long as the outcome is registered as certain macroscopic configuration of particles \citep{bell2004speakable}. We can think of the new law in $T_{M2}$ as akin to the Hamiltonian function in classical mechanics, which is interpreted as encoding all the classical force laws, except that specifying $F(x,t)$ is much more complicated than specifying a typical Hamiltonian. Both $F(x,t)$ and the Hamiltonian are components of respective laws of nature that tell particles how to move.\footnote{Note that we can  decompose the standard ontology in many other dimensions, corresponding to more ways to generate empirically equivalent laws for a Maxwellian world. This move is discussed at length by \cite{AlbertLPT}. Similar strategies have been considered in the ``quantum Humeanism'' literature. See \cite{miller2013quantum}, \cite{esfeld2014quantum}, \cite{callender2015one, bhogal2015humean}, and \cite{chen2018HU}.} Given metaphysical realism, at most one of the two theories has the correct nomology.\\

\noindent
\textbf{Algorithm B: Changing the nomology directly}.  
  
 \textit{  General strategy.}  This strategy is designed for MinP.  We can generate empirical equivalence by directly changing the nomology. Suppose the actual mosaic $\xi$ is governed by the  law $L_1$. Consider $L_2$, where $\Omega^{L_1} \neq \Omega^{L_2}$ and $\xi \in \Omega^{L_2}$.  $L_1$ and $L_2$ are distinct laws because they have distinct sets of models. Since $E$ (which can be regarded as a coarse-grained and partial description of $\xi$) can arise from both, the two laws are empirically equivalent.  There are infinitely many such candidates for $\Omega^{L_2}$. For example, $\Omega^{L_2}$ can be obtained by replacing one mosaic in $\Omega^{L_1}$ with something different and not already a member of $\Omega^{L_1}$, by adding some mosaics to $\Omega^{L_1}$, or by removing some mosaics in $\Omega^{L_1}$.  $L_2$ is empirically equivalent with $L_1$ since $E$ is compatible with both.\footnote{See \cite{manchak2009can, manchak2020global} for more examples.}
 
 \textit{Example.} Let $L_1$ be the Einstein equation of general relativity, with  $\Omega^{L_1}=\Omega^{GR}$, the set of general relativistic spacetimes. Assume that the actual spacetime is governed by $L_1$, so that $\xi\in \Omega^{L_1}$. Consider $L_2$, a law that permits only the actual spacetime and completely specifies its microscopic detail, with $\Omega^{L_2}=\{\xi\}.$ Since our evidence $E$ arises from $\xi$, it is compatible with both $L_1$ and $L_2$. Since it needs to encode the exact detail of $\xi$, in general $L_2$ is much more complicated than $L_1$.\footnote{$L_2$ is a case of strong determinism. See \cite{adlam2021determinism} and \cite{chen2022strong} for recent discussions.} Given metaphysical realism, at most one of $L_1$ and $L_2$ corresponds to the actual law.\\
  
  \noindent
\textbf{Algorithm C: Changing the nomology by changing the ontology}.  
  
 \textit{  General strategy.}  This strategy is designed for BSA. On BSA, we can change the nomology by making suitable changes in the ontology (mosaic), which will in general change what the best system is. Suppose the actual mosaic $\xi$ is optimally described by the actual best system $L_1=BS(\xi)$. We can consider a slightly different mosaic   $\xi'$, such that it differs from $\xi$ in some spatiotemporal region that is never observed and yet $E$ is compatible with both $\xi$ and $\xi'$. There are infinitely many such candidates for $\xi'$ whose best system $L_2=BS(\xi')$ differs from $L_1$. Alternatively, we can expand $\xi$ to  $\xi' \neq \xi$ such that $\xi$ is a proper part of $\xi'$. There are many such candidates for $\xi'$ whose best system $L_2=BS(\xi')$ differs from $L_1$, even though $E$ is compatible with all of them.

\textit{Example.} Let $L_1$ be the Einstein equation of general relativity, with  $\Omega^{L_1}=\Omega^{GR}$, the set of general relativistic spacetimes. Assume that the actual spacetime is globally hyperbolic and optimally described by $L_1$, so that $L_1 = BS(\xi)$. Consider $\xi'$, which differs from $\xi$ in only the number of particles in a small spacetime region $R$ in a far away galaxy that no direct observation is ever made. Since the number of particles is an invariant property of general relativity, it is left unchanged after a ``hole transformation'' \citep{sep-spacetime-holearg}.  We can use determinism to deduce that $\xi'$ is incompatible with general relativity, so that $L_1 \neq BS(\xi')$. Let $L_2$ denote $BS(\xi')$. $L_1\neq L_2$ and yet they are compatible with the same evidence we obtain in $\xi$.  Since $\xi'$ violates the conservation of number of particles, $L_2$ should be more complicated than $L_1$. Given metaphysical realism, at most one of $L_1$ and $L_2$ corresponds to the actual law.

 Algorithms A, B, and C can be combined to generate more sophisticated cases of empirical equivalence. The question they raise is this: what breaks the tie among empirical equivalents and epistemically justifies our belief in the intuitively correct  law, given the evidence that we will in fact obtain? In other words, in cases of empirical equivalence, how can we hold on to epistemic realism given our commitment to metaphysical realism? To do so, we need a tie breaker. 

\subsection{Nomic Simplicity}

 I propose that \textit{simplicity is a fundamental epistemic guide to lawhood}. Roughly speaking, simpler candidates are more likely to be laws, all else being equal.   It  secures epistemic realism in cases of empirical equivalence where simplicity is the deciding factor. In particular, we should accept this principle:
\begin{description}
  \item[Principle of Nomic Simplicity (PNS)] Other things being equal, simpler propositions are more likely to be laws. More precisely, other things being equal, for two propositions $L_1$ and $L_2$, if $L_1>_S L_2$, then $L[L_1] >_P L[L_2]$, where $>_S$ represents the comparative simplicity relation, $>_P$ represents the comparative probability relation, and $L[\cdot]$ denotes \textit{is a law}, which is an operator that maps a proposition to one about lawhood.\footnote{For example, $L[F=ma]$ expresses the proposition that \textit{F=ma is a law}. The proposition \textit{F=ma} is what \cite{lange2009laws} calls a ``sub-nomic proposition.'' } 
\end{description}
 As a fundamental principle, PNS is not justified by anything else.\footnote{Saying that simplicity is a fundamental epistemic guide to lawhood does not mean it is the only such guide. Recall that  PNS contains the proviso ``other things being equal.'' But sometimes other factors are not held equal, and we need to consider overall comparisons of theoretical virtues (epistemic guides) and their balance. Other theoretical virtues can also serve as epistemic guides for lawhood. For example, informativeness and naturalness are two such virtues. A simple equation that does not describe much or describe things in too gruesome manners is less likely to be a law. We can formulate a more general principle: 
\begin{description}
  \item[Principle of Nomic Virtues (PNV)] 
  For two propositions $L_1$ and $L_2$, if $L_1>_O L_2$, then $L[L_1] >_P L[L_2]$, where $>_O$ represents the relation of overall comparison that takes into account all the theoretical virtues and their tradeoffs, of which of which $>_S$ is a contributing factor,  $>_P$ represents the comparative probability relation, and $L[\cdot]$ denotes \textit{is a law}, which is an operator that maps a proposition to one about lawhood.
\end{description}
What is overall better is a holistic matter, and it can involve trade-offs among the theoretical virtues such as simplicity, informativeness, and naturalness. PNV should be thought of as the more general epistemic principle than PNS. I shall mainly focus on PNS, but what I say below should carry over to PNV.}  But what is special about PNS, and why not use the oft-cited principle of simplicity as below?
\begin{description}
  \item[Principle of Simplicity (PS)] Other things being equal, simpler propositions are more likely to be true. More precisely, other things being equal, for two propositions $L_1$ and $L_2$, if $L_1>_S L_2$, then $L_1 >_P L_2$, where $>_S$ represents the comparative simplicity relation, $>_P$ represents the comparative probability relation.\footnote{It may be too demanding to require a total order that induces a \textit{normalizable} probability distribution over the space of all possible laws.  It is less demanding to formulate PS in terms of comparative probability.}
\end{description}
The reason is that PS leads to probabilistic incoherence while PNS straightforwardly avoids it. Whenever two theories have nested sets of models, say $\Omega^{L_1} \subset \Omega^{L_2}$,  the probability that $L_1$ is true cannot be higher than the probability that $L_2$ is true. For concreteness, consider an example from spacetime physics. Let $\Omega^{GR}$ denote the set of models compatible with the fundamental law in general relativity---the Einstein equation, and let $\Omega^{GR^+}$ denote the union of  $\Omega^{GR}$ and a few random spacetime models that do not satisfy the Einstein equation (see Figure 7). Suppose there is no simple law that generates $\Omega^{GR^+}$. While the law of $GR$ (the Einstein equation) is presumably simpler than that of $GR^+$, the former cannot be more likely to be true than the latter, since every model of  $GR$ is a model of $GR^+$, and not every model of $GR^+$ is a model of $GR$. This is an instance of the problem of nested theories, as $\Omega^{GR}$ is a subclass of and nested within $\Omega^{GR^+}$. 

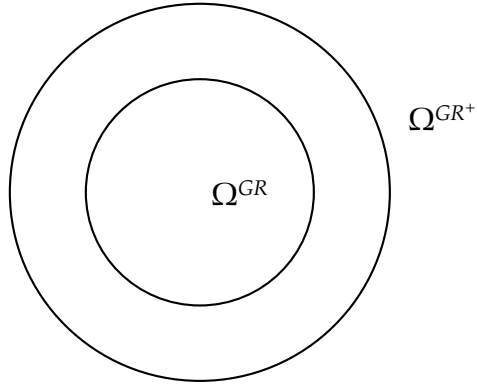
\begin{figure}
\centering
\begin{tikzpicture}
\coordinate (O) at (0,0);
\draw[thick] (O) circle (2.5);
\draw[thick] (O) circle (1.5) node[right] {$\Omega^{GR}$};
  \fill (2.6, 1) circle (0pt) node[right] {$\Omega^{GR^+}$};
\end{tikzpicture}
\caption{$\Omega^{GR}$ is nested within $\Omega^{GR^+}$.}
\end{figure}

 On PS, in the case of nested theories, we have probabilistic incoherence. If $L_1$ is simpler than $L_2$, applying the principle that simpler laws are more likely to be true, we have $L_1 >_P L_2$. However, if $L_1$ and $L_2$ are nested with $\Omega^{L_1} \subset \Omega^{L_2}$,  the axioms of probability entail that  $L_1 \leq_P L_2$. Contradiction! 

On PNS, we can avoid the problem. Even though we think that the Einstein equation is more likely to be a law, it is less likely to be true than the law of $GR^+$.  According to PNS,  what simplicity selects here is not truth in general, but truth about lawhood, i.e. whether a certain proposition has the property of being a fundamental law. Let us assume that fundamental lawhood is factive, which is granted on both BSA and MinP. Hence, lawhood implies truth:  $L[p] \Rightarrow p$. However, truth does not imply lawhood:   $p \nRightarrow L[p]$.  This shows that $L[p]$ is logically inequivalent to $p$. This is the key to solve the problem of coherence. 

 On PNS, the contradiction that inflicts PS is removed, because \textit{more likely to be a law} does not entail \textit{more likely to be true}.  If $L_1$ and $L_2$ are nested, where $L_1$ is simpler than $L_2$ but  $\Omega^{L_1} \subset \Omega^{L_2}$, then $L_1 \leq_P L_2$. It is compatible with the fact that  $L[L_1] >_P L[L_2]$. What we have is an inequality chain: 
 \begin{equation}
 L[L_2] <_P L[L_1] \leq_P L_1 \leq_P L_2
\end{equation}
From the perspective of nomic realism, one can consistently endorse PNS without endorsing PS. Some facts are laws, but not all facts are laws. Laws correspond to a special set of facts. On BSA, they are the best-system axioms. On MinP, they are the constraints on what is physically possible. 

This is also a new and simple solution to the problem of nested theories / problem of coherence. It is compatible with but less demanding and perhaps more general than the recent proposal of \cite{henderson2022inference}. Unlike Henderson's approach, my proposal works even when candidate laws are not structured in a hierarchy.

\subsection{Theoretical Benefits}
We should accept PNS because of its theoretical benefits. Here I discuss six attractive consequences of PNS. 

\textit{Empirical equivalence.} PNS is useful for resolving cases of empirical equivalence constructed along Algorithms A-C. For Algorithm A, $T_2$ will in general employ much more complicated laws than $T_1$. For example,  the laws of $T_{M2}$ specify $F(x,t)$ in its exact detail. For Algorithm B, $L2$ will in general be more complicated than $L1$, if $\Omega^{L_2}$ is obtained from $\Omega^{L_1}$ by adding or subtracting a few models. For example, a strongly deterministic theory of some sufficiently complex general relativistic spacetime, as described in the example, needs to specify the exact detail of that spacetime, will employ laws much more complicated than the Einstein equation. For Algorithm C, even though the mosaics of $L1$ and $L2$ are not that different, if $L1$ is a simple system, then  in general $L2$ will not be. In fact, given enough changes from the actual mosaic, there may not be any optimal system that simplifies the altered mosaic to produce a good system. With PNS, we are justified in choosing the theoretical package with simpler laws, which agrees with standard theory choice.

\textit{Induction.} We want to know what the physical reality $(L, \xi)$ is like. Given our limited evidence about some part of $\xi$ and some aspect of $L$, what justifies our inference to other parts of $\xi$ or other aspect of $L$ that will be revealed in upcoming observations or in observations that could have been performed? It does not follow logically. Without some \textit{a priori} rational guide to what $(L, \xi)$ is like or probably like, we have no rational justification for favoring $(L, \xi)$ over any alternative compatible with our limited evidence. On a given $L$ we know what kind of $\xi$ to expect. But we are given neither $L$ or $\xi$.  Without further inferential principles, it is hard to make sense of the viability of induction.

In \S2.6, we discussed the potential connection between laws and induction.  What does induction require from laws? One might demand that laws  hold the same way everywhere and everywhen, but that is either vacuous or too inflexible. A better answer is the simplicity of  laws, as required by PNS. A simple law can give rise to a complicated mosaic with an intricate matter distribution. The complexity is only apparent, because behind all the surface phenomena is a simple law that is discoverable. By assuming that the universe is governed by a simple law, one may make reasonable guesses about unobserved parts of the universe, based on a simple rule compatible with observed data. But nomic simplicity does not forbid laws about boundary conditions or about particular individuals.  Some simple laws may even have temporal variations, such as a time-dependent function $F=\frac{1}{t+k}ma$ for some constant $k$.  As long as the temporal variation and spatial variation are not too extreme as to require complicated laws, we can still inductively learn about physical reality based on available evidence, even in a non-uniform spacetime with dramatically different events in different regions.

\textit{Symmetry.} Symmetry principles play important roles in theory construction and discovery. Physicists routinely use symmetries to justify or guide their physical postulates. However, whether symmetries hold is an empirical fact, not guaranteed by the world \textit{a priori}. So why should we regard symmetry principles to be useful, and what are they targeting? I suggest that certain applications of symmetry principles are defeasible guides for finding simple laws. In such cases, their epistemic usefulness may come from the fact that certain symmetries are defeasible indicators for simplicity.\footnote{For a related perspective, see \cite{north2021physics}.}  Consider  the toy example
\begin{equation}\label{E1}
  F=ma \text{ for } (-\infty, t] \text{ and } F=(8m^9-\frac{1}{7}m^5+\pi m^3+km^2+m)a \text{ for } (t, \infty)
\end{equation}
with $F$ given by Newtonian gravitation. This law violates time-translation invariance and time-reversal invariance. In this case, we have a much better law that is time-translation and time-reversal invariant: 
 \begin{equation}\label{E2}
  F=ma \text{ for all times}  
\end{equation}
The presence of the two symmetries in (\ref{E2}) and the lack of them in (\ref{E1}), indicate that all else being equal we should prefer  (\ref{E2}) to  (\ref{E1}). We can explain this preference by appealing to their relative complexity. (\ref{E2}) is much simpler than (\ref{E1}), and the existence of the symmetries are good indicators of the relative simplicity.  However, in this comparison, we are assuming that both equations are valid for the relevant evidence (evidence obtained so far or total evidence that will ever be obtained). The preference is compatible with the fact that \textit{if} empirical data is better captured by (\ref{E1}), we should prefer (\ref{E1}) to (\ref{E2}).

In the relevant situations where symmetry principles are guides to simplicity, they are only defeasible guides.  Symmetry principles are not an end in itself for theory choice.  I shall provide two more examples to show that familiar symmetry principles are not sacred, but rather defeasible indicators of simplicity, that can be ultimately sacrificed if we already have a reasonably simple theory that is better than the alternatives. 

The first is the toy example of the Mandelbrot world (\S3.2). The physical reality consisting of $(L_M, \xi_M)$ is friendly to scientific discovery. If we were inhabitants in that world, we can  learn the structure of the whole $\xi_M$ from the structure of its parts, by learning what $L_M$ is. However, $L_M$ is not a law with any recognizable spatial or temporal symmetries.\footnote{There is, however, the reflection symmetry about the real axis. But it does not play any useful role here, and we can just focus on the upper half of the Mandelbrot world if needed.} Nevertheless, the physical reality described by $(L_M, \xi_M)$ is a perfect example of an ultimate theory (though not of the actual world). It is an elegant and powerful explanation for the patterns in the Mandelbrot world. What could be a better explanation? I suggest that none would be, even if it had more symmetries. In this case, we do not need symmetry principles to choose the right law, because we already have a simple and good candidate law.  The lack of symmetries is not a regrettable feature of the world, but a consequence of its simple law. 

The second and more realistic example is the Bohmian Wentaculus \citep{chen2018IPH, chen2018HU, chen2023detlef}. With the Initial Projection Hypothesis (\S3.3.2), the initial quantum state is as simple as the Past Hypothesis. This allows us to adopt the nomic interpretation of the quantum state, and understand the mosaic $\xi_B$ as consisting of only particle trajectories in spacetime, with the fundamental dynamical law $L_B$  as given by this differential equation: 
 \begin{equation}\label{N4}
\frac{dQ_i}{dt} = 
\frac{\hbar}{m_i} \text{Im} \frac{\nabla_{q_{i}}   W_{IPH} (q, q', t)}{W_{IPH} (q, q', t)} (Q) = 
\frac{\hbar}{m_i} \text{Im} \frac{\nabla_{q_{i}}   \bra{q} e^{-i \hat{H} t/\hbar} \hat{W}_{IPH} ( t_0) e^{i \hat{H} t/\hbar} \ket{q'} }{ \bra{q} e^{-i \hat{H} t/\hbar} \hat{W}_{IPH} (t_0) e^{i \hat{H} t/\hbar} \ket{q'}} (q=q'=Q)
\end{equation}
Since the quantum state is nomic, as specified by a law, the right hand side should be the canonical formulation of the fundamental dynamical law for this world. Notice that the right hand side of the equation is not time-translation invariant, as at different times the expression 
$$\text{Im} \frac{\nabla_{q_{i}}   \bra{q} e^{-i \hat{H} t/\hbar} \hat{W}_{IPH} ( t_0) e^{i \hat{H} t/\hbar} \ket{q'} }{ \bra{q} e^{-i \hat{H} t/\hbar} \hat{W}_{IPH} (t_0) e^{i \hat{H} t/\hbar} \ket{q'}}$$ 
will in general take on different forms. However, the physical reality described by the Bohmian Wentaculus may be our world, and the equation can be discovered scientifically. The law is a version of the Bohmian guidance equation that directly incorporates a version of the Past Hypothesis. Hence, $(L_B, \xi_B)$ describes a physical reality that is friendly to scientific discovery and yet does not validate time-translation invariance. 

In the Bohmian Wentaculus world, symmetry principles can be applied, but the fundamental dynamical law explicitly violates time-translation invariance. In such cases, the lack of symmetries is not a problem, because we already have found the simple candidate that has the desirable features. Again, the time-translation non-invariance is a consequence of its simple law. PNS takes precedence over symmetry principles and are the deeper justification for theory choice.

\textit{Dynamics.} We have good reasons to allow fundamental laws of boundary conditions. However, many boundary conditions are not suitable candidates for fundamental lawhood. Epistemic guides such as simplicity allow us to be selective in postulating boundary condition laws, and to give more weight to proposals that include dynamical laws. 

 The examples of good boundary condition laws have a common feature: they are simple to specify. Many boundary conditions contain a great deal of correlations, but only a select few are good candidates for fundamental laws, namely those that are also sufficiently simple. One may wonder why we choose the Past Hypothesis, a macroscopic description,  over a precise microscopic initial condition of the universe. The answer is that the former is much simpler than the latter and is still sufficiently powerful to explain a variety of temporally asymmetric regularities. The simplicity of the boundary condition laws  make it almost inevitable that we will have dynamical laws in addition to boundary condition laws. The scientific explanations of natural phenomena come from the combination of simple boundary conditions and dynamical laws. As such, dynamical laws have to carry a lot of information by themselves.

\textit{Determinism.} Nomic realism is often accompanied with other reasonable expectations about  laws. 
  On MinP, given any mosaic $\xi$, there are many possible choices of $L$ such that $\xi \in \Omega^L$ and mosaics do not cross in $\Omega^L$. Here is an algorithm to generate some of them: construct a two-member set $\Omega^L=\{\alpha, \beta \}$ such that $\alpha$ and $\beta$ agree at no time (or any Cauchy surface). Any law with such a domain meets the definition of determinism. As long as $\alpha$ is not a world where every logically possible state of the universe happens some time in the universe, it is plausible to think that there are many different choices of $\beta$ that can ensure determinism. Without a further principle about what we should expect of $L$, determinism is too easy and almost trivial on MinP.   On BSA, the problem is the opposite. It becomes too difficult and almost impossible for a world to be deterministic. Given the evidence $E$ we have about the mosaic, even though $E$ may be optimally summarized by a deterministic law $L$, it does not guarantee (or make likely, without further assumptions) that the entire mosaic is optimally summarized by a deterministic law $L$. Small ``perturbations'' somewhere in the mosaic can easily make its best system fail determinism.\footnote{See \cite{builes2022ineffability} for a related argument.}

 Hence, there is a question of what nomic realists should say that constitutes a principled reason to think that determinism is not completely trivial (on MinP) and not epistemically inaccessible (on BSA). 
  With PNS, determinism is no longer trivial on MinP. Given any mosaic $\xi$, even though there are many deterministic candidates compatible with and true at $\xi$. Not every mosaic will be compatible with a relatively simple law that is deterministic. The non-triviality of determinism on MinP is the fact that it is non-trivial to find a law that is simple and deterministic, as that is not guaranteed for every metaphysically possible mosaic. 
With PNS, determinism is no longer epistemically inaccessible on BSA. This follows from the more general principle that PNS gives us epistemic justification to hold beliefs about parts of the mosaic that we have not observed and will never observe. We are justified in believing that the best system of the actual mosaic is relatively simple, even though the actual evidence does not entail that. If the actual evidence can be optimally summarized by a deterministic law restricted to the actual evidence, we have epistemic justification to make inferences about regions that will not be observed -- the entire mosaic, $\xi$, can be summarized by a simple law that happens to be deterministic.

\textit{Explanation.} There is a strong connection between nomic realism and scientific explanation. The point of postulating  laws, on BSA and on MinP, is to provide scientific explanations. However, not all candidate laws provide the same quality of explanation or same kind of explanation. Hence, on both versions of nomic realism, we might wonder if there is a principled reason to think that we will have a successful scientific explanation for all phenomena. 

On MinP, explanations must relate to us (\S3.2). Constraints, in and of themselves, do not always provide satisfying explanations.  Many constraints are complicated and thus insufficient for understanding nature.  For example, the constraint given by just $\Omega^L = \{\xi_M\}$, which requires a complete specification of the mosaic, is insufficient for understanding the Mandelbrot world. Knowing why there is a pattern requires more than knowing the exact distribution of matter.   On MinP, many candidate laws can constrain the mosaic. But not all have the level of simplicity to provide illumination about the mosaic. With PNS, we expect the actual constraint to be relatively simple. The constraint given by the Mandelbrot law should be preferred to that  given by $\Omega^L = \{\xi_M\}$. The simple law provides a successful explanation while the more complicated one does not. 

On BSA, it is built in the notion of laws that they systematize the mosaic. However, whether there is a systematization that is simpler than the mosaic is a contingent matter, depending on the detailed, microscopic, and global structures of the mosaic. Not every mosaic supports a systematization that provides illumination in the sense of unifying the diverse phenomena in the mosaic. BSA only entails that the best system is no more complex than the exact specification of the mosaic. For example, some mosaics may support no better optimal summary than the exact specification of the mosaic itself. Hence, on BSA, having successful explanations is not automatic. It requires the mosaic to be favorable.  On BSA, some mosaics are favorable: they support optimal summaries that are simpler than themselves and provide ``Humean explanations'' about the mosaic. In fact, in a sense, most mosaics are not favorable \citep{lazarovici2020typical}.  There exist mosaics underdetermined by actual evidence that do not support any good summaries. Given the actual evidence, with PNS, we are epistemically justified in inferring that the actual best system is relatively simple such that it can provide a ``Humean explanation'' about the actual mosaic. In effect, we are expecting that the actual Humean mosaic is a favorable one that completely cooperates with our scientific methodology and is such that it can be unified in a reasonably simple best system.

\subsection{Summary}
Nomic realists should postulate the principle of nomic simplicity. It vindicates epistemic realism when there is empirical equivalence (at least in those cases discussed in the paper), avoids probabilistic incoherence when there are nested theories, and supports realist commitments regarding induction, symmetries, dynamics, determinism, and explanation. With many theoretical benefits for only a small price, it is a great bargain.

\section{Exactness}

Another hallmark of  laws is their exactness, in contrast to the pervasive vagueness we find in ordinary language. A good way of understanding something is to study its opposite. In this section, I discuss a model of nomic vagueness and a case study. 

\subsection{Nomic Exactness}
 Many predicates we use in everyday contexts do not have determinate boundaries of application.  Is John bald when he has exactly 5250 hairs on his head? There are determinate cases of ``bald,'' but there are also  borderline cases of ``bald.'' In other words,  predicates such as ``bald'' are indeterminate: there are individuals  such that it is indeterminate whether they are bald.\footnote{There are subtleties  about how best to characterize vagueness.  For  reviews on vagueness and the sorites, see  \citealt{KeefeSmith1}, \citealt{sep-vagueness}, and \citealt{sep-sorites-paradox}. } Moreover, the boundaries between  ``bald'' and ``borderline bald'' are also indeterminate. There do not seem to be sharp boundaries \emph{anywhere}, a phenomenon as higher-order vagueness.  Vagueness gives rise to many paradoxes (such as the sorites) and serious challenges to classical logic. 

\x{We might expect} that, at the level of fundamental physics, the kind of vagueness that ``infects'' ordinary language should disappear. That is, the fundamental laws of physics, the predicates they invoke,  and the properties they refer to should be exact. The expectation is supported  by the history of physics and  the ideal that physics should deliver an objective and precise description of nature. All the paradigm cases of candidate fundamental laws of nature are not only simple and universal, but also \emph{exact}, in the sense that, for every class of worlds (or class of solutions), fundamental laws either determinately apply or determinately fail. Suppose the fundamental laws are Newton's equation of motion $F=ma$ and law of universal gravitation $F=Gm_1m_2/r^2$: there is no vagueness about whether a physical history (described in terms of  trajectories of point particles with Newtonian masses) satisfies the laws. In other words, nomologically possible worlds form a set. 

Fundamental nomic exactness---the ideal (roughly) that fundamental laws are exact---supports an important principle about the mathematical expressibility of fundamental laws. If some fundamental laws are vague, it will be difficult to describe them mathematically in a way that genuinely respects  their vagueness  and does not impose  sharp boundaries anywhere. The kind of mathematics we are used to, built from a set-theoretic foundation, does not lend itself naturally to model the genuine fuzziness of vagueness. One could go further: the language of mathematics and the language of fundamental physics are supposed to be exemplars for  the ``ideal language,'' a language that is exact, suggested in Frege's \emph{Begriffsschrift},  Russell's logical atomism, and Leibniz's \emph{characteristica universalis}. The successful application of mathematical equations in formulating  laws \textit{seems} to leave no room  for vagueness to enter  into a fundamental physical theory. If there is fundamental nomic vagueness, and if vagueness is not completely mathematically expressible, then the fundamental physical theory is not completely mathematically expressible.

\subsection{Nomic Vagueness}

\begin{figure}
\centerline{\includegraphics[scale=0.2]{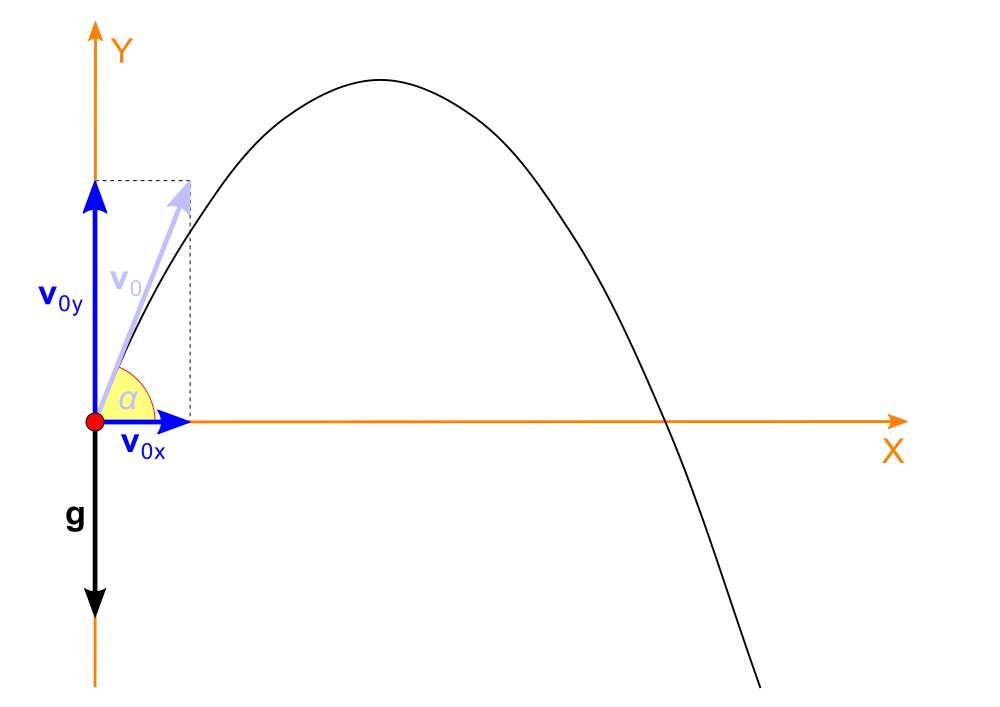}}
\caption{The motion of a projectile in Newtonian mechanics. Picture from Zátonyi Sándor, (ifj.) Fizped, CC BY-SA 3.0 <https://creativecommons.org/licenses/by-sa/3.0>, via Wikimedia Commons} 
\end{figure}

How should we understand the exactness of  paradigm fundamental laws of nature? Let us start with  the familiar case of Newtonian mechanics (with Newtonian gravitation). Its laws can be expressed as a set of differential equations that admit a determinate set of solutions. Those solutions specify all and only the possible histories compatible with the laws; each solution  represents a nomologically possible world  of the theory. 

Consider the projectile motion illustrated in Figure 8. Suppose that the projectile has unit mass $m$ and the gravitational acceleration is  $g$. We can specify the history of the projectile with the initial height, initial velocity, maximum height, and distance traveled. There is a set of histories compatible with  the laws. For any history of the projectile, it is either determinately compatible with the equations or determinately incompatible with the equations.  

If $W$ represents the space of all possible worlds, then the nomologically possible worlds of Newtonian mechanics corresponds to $\Omega^{NM}$, a proper subset in $W$ that has a determinate boundary, \x{where the boundary is not in spacetime but in modal space.} For any possible world $w \in W$, either $w$  is  contained in $\Omega^{NM}$ or it is not. For example, in Figure 9, $w1$ is inside but $w2$ is outside $\Omega^N$. In other words, $w1$ is nomologically possible while $w2$ is nomologically impossible if Newtonian laws are true and fundamental. Call $\Omega^{NM}$ the \textit{domain} of Newtonian mechanics. 
 We can capture an aspect of  \x{fundamental} nomic exactness as the exactness of the domain: 

\begin{description}
	\item[Domain Exactness] \x{A fundamental law $L$ is domain-exact if and only if, (a) for any world $w \in W$, there is a determinate fact about whether $w$ is contained inside $L$'s domain of worlds, i.e. $L$'s domain has no borderline worlds, (b) $L$'s domain, which may also be called $L$'s \textit{extension}, forms a set of worlds,   (c) $L$'s domain is not susceptible to sorites paradoxes, and (d) $L$'s domain has no borderline borderline worlds, no borderline borderline borderline worlds, and so on.}  
\end{description}

\begin{figure}
\centerline{\includegraphics[scale=0.22]{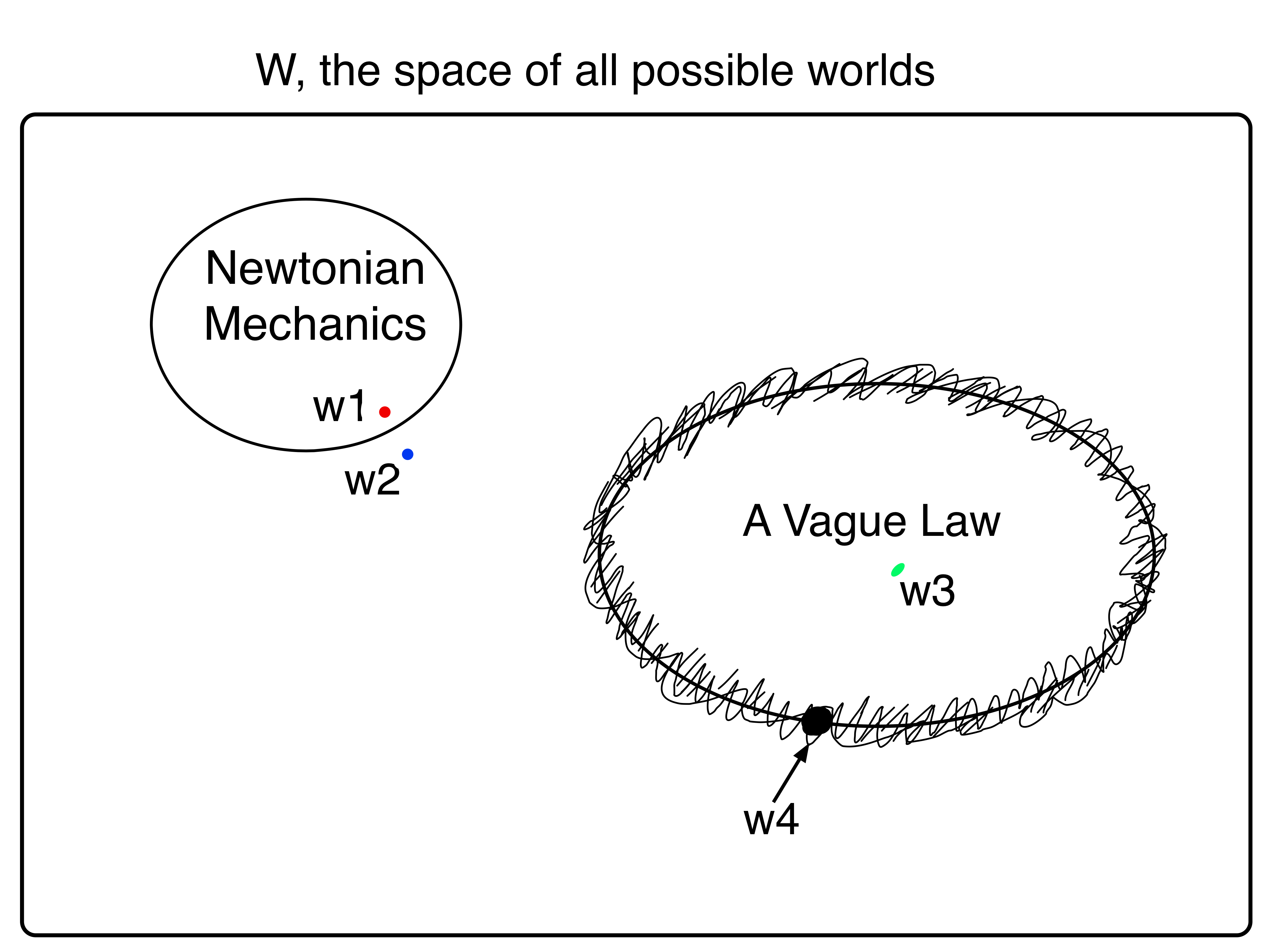}}
\caption{An exact fundamental law and a vague fundamental law represented in modal space. }
\end{figure}

\x{In contrast, a domain-vague law has none of (a)--(d). Intuitively, a domain-vague law has a vague boundary in the following sense.} In Figure 9, a domain-vague law is pictured by a ``collection'' of worlds with a fuzzy boundary. Just as a cloud does not have a clear starting point or a clear end point, the fuzzy ``collection'' of worlds does not delineate the worlds into those that are clearly compatible and those that are clearly incompatible with the law. To borrow the words of \cite{sainsbury1990concepts}, a domain-vague law classifies  worlds ``without setting boundaries'' in modal space. For example, $w3$ is clearly contained inside the domain of the vague law, since it is so far away from the fuzzy boundary; but $w4$ is not clearly contained inside the domain of the vague law, and neither is it clearly outside; $w2$ is clearly outside the domain. \x{More precisely, I propose that we understand domain vagueness as the opposite of domain exactness: }

\begin{description}
	\item[Domain Vagueness] \x{A fundamental law $L$ is domain-vague if and only if $L$ meets all four conditions below.}  
\end{description}
\begin{itemize}
	\item[ (a')] $L$ has borderline worlds that are not determinately compatible with it. For some world $w\in W$, there fails to be a determinate fact about whether $w$ is contained inside $L$'s domain of worlds. 
	\item[(b')] $L$  lacks a well-defined extension in terms of a set of models or a set of nomological possibilities. Nomological necessities and possibilities turn out to be vague. 
	\item[(c')] $L$ is susceptible to sorites paradoxes. We can start from a world that is determinately lawful, proceed to gradually make small changes to the world along some relevant dimension, and eventually arrive at a world that is determinately unlawful. But no particular small change makes the difference between determinately lawful and determinately unlawful. 
	\item[(d')] $L$ possesses higher-order domain-vagueness. Whenever there are borderline lawful worlds, there are borderline borderline lawful worlds, and so on. It seems inappropriate to draw a sharp line anywhere. This reflects the genuine fuzziness of domain vagueness.
\end{itemize}
Domain vagueness has features similar to those of ordinary-language vagueness. 
Domain exactness and domain vagueness capture the kind of  \x{fundamental} nomic exactness and  \x{fundamental}  nomic vagueness we are most interested in here. (There is another kind of  \x{fundamental}  nomic vagueness that results from vague objective probabilities or typicalities. See \cite{goldstein2012typicality}. \cite{fenton2019imprecise} offers an account of imprecise (but not vague) chances in the best-system theory.)

\subsection{A Case Study}

The Past Hypothesis (PH) presents a case of fundamental nomic vagueness. To begin, let us consider the following version of PH that is sometimes proposed:

\begin{description}
	\item[Super Weak Past Hypothesis (SWPH)] At one temporal boundary of space-time, the universe has very low entropy. 
\end{description}
SWPH is obviously vague. How low is low? The collection of worlds with ``low-entropy'' initial conditions has fuzzy boundaries in the space of possible worlds. Hence, if SWPH were a fundamental law, then we would have domain vagueness.  

However, SWPH may not be  detailed enough to explain all the temporal asymmetries. For example, in order to explain the temporal asymmetries of records, intervention, and knowledge, \cite{albert2000time} and \cite{loewer2016mentaculus} suggest that we need a more specific condition that narrows down the initial microstates to a particular macrostate. One way to specify the macrostate invokes exact numeral values for the macroscopic variables of the early universe. Let $S_0, T_0, V_0, D_0$ represent the exact values (or exact distributions) of (low) entropy, (high) temperature, (small) volume, and (roughly uniform) density distribution of the initial state. Consider the following version of PH:  
\begin{description}
	\item[Weak Past Hypothesis (WPH)] At one temporal boundary of space-time, the universe is in a particular macrostate $M_0$, specified by the macroscopic variables $S_0, T_0, V_0$, and $D_0$. 
\end{description}
WPH is a stronger version of PH than SWPH. By picking out a particular (low-entropy) macrostate $M_0$ from many macrostates,  WPH more severely constrains the initial state of the universe. WPH is also more precise than SWPH. (Some may  even complain that the WPH is too strong and too precise.) Unfortunately, WPH is still vague. The collection of worlds compatible with WPH  has fuzzy boundaries. If WPH were a fundamental law, then we would still have nomic (domain) vagueness: there are some worlds whose initial conditions are borderline cases of being in the macrostate $M_0$, specified by the macroscopic variables $S_0, T_0, V_0$, and $D_0$.

The vagueness of WPH is  revealed when we  connect the macroscopic variables to the microscopic ones. Which set of microstates realizes the macrostate $M_0$? There is hardly any sharp boundary between those that do and those that do not realize the macrostate. A macrostate, after all, is a coarse-grained description of the physical state. As with many cases of coarse-graining, there  can be borderline cases. (To connect to our discussion in \S2.1, the vagueness of macrostates is similar to the vagueness of ``is bald'' and ``is a table.'') In fact, a macrostate can be vague even when it is specified with precise values of the macro-variables. This point should be familiar to those working in the foundations of statistical mechanics.\footnote{Commenting on the vagueness of the macrostate boundaries, \cite{LoewerCatSLaw} writes, ``Obviously, the notion of \emph{macro state} is vague and there are many precisifications that would serve the purposes of statistical mechanics.''  \cite{goldstein2019gibbs} write, ``there is some arbitrariness in where exactly to `draw the boundaries.'''} However, it is worth spelling out the reasons to understand where and why such vagueness exists.

 \begin{figure}
\centerline{\includegraphics[scale=0.24]{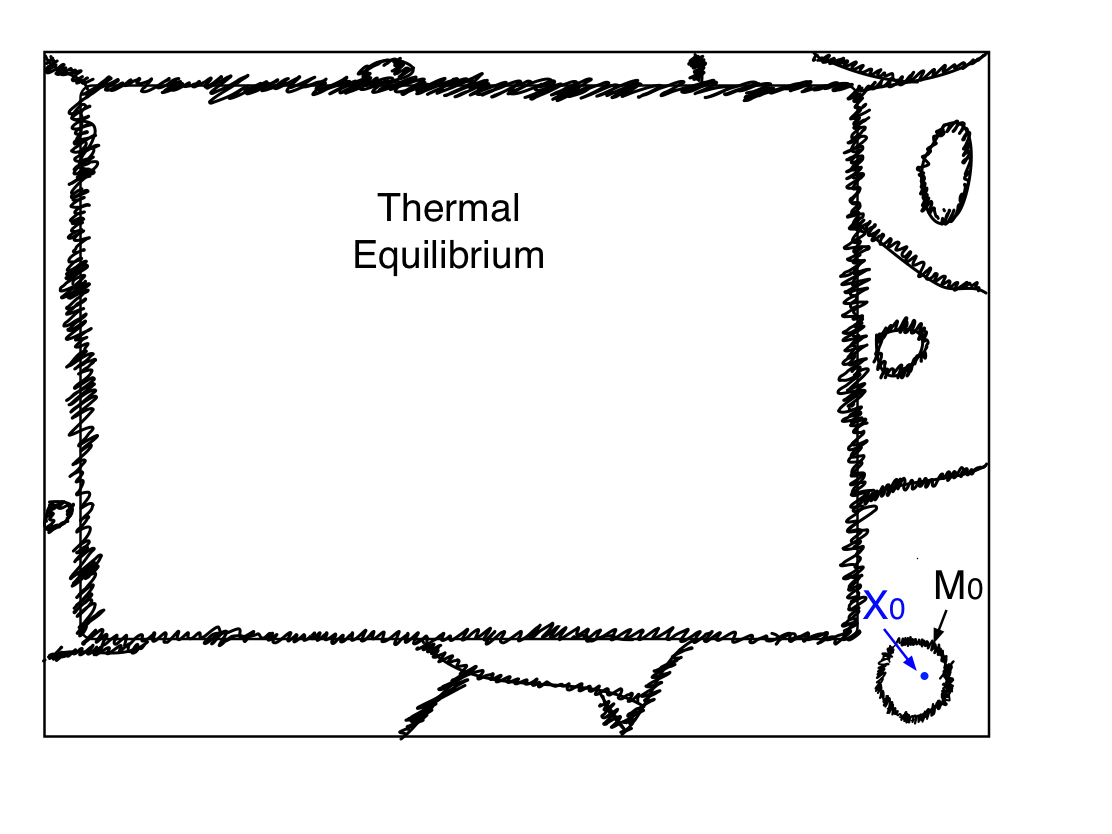}}
\caption{A diagram of  phase space where macrostates have fuzzy boundaries. The macrostate $M_0$ represents the initial low-entropy condition described by WPH. $X_0$ is the actual initial microstate. The picture is not drawn to scale. }
\end{figure}

There is a systematic way to think about the vagueness of the thermodynamic macrostates in general and the vagueness of $M_0$ in the WPH. In the Boltzmannian account of  classical statistical mechanics, macrostates and microstates can be understood as certain structures on  phase space (Figure 10). 
\begin{itemize}
	\item Phase space: in classical mechanics, phase space is a 6N-dimensional space that encodes all the microscopic possibilities of the system. 
	\item Microstate: a point in phase space, which is a maximally specific description of a system. In classical mechanics, the microstate specifies the positions and the momenta of all particles.  
	\item Macrostate: a region in phase space in which the points inside are macroscopically similar, which is a less detailed and more coarse-grained description of a system. The largest macrostate is thermal equilibrium.
	\item Fuzziness: the partition of phase space into macrostates is not exact; the macrostates have fuzzy boundaries. Their boundaries  become exact only given some choices of the ``C-parameters'', including the size of  cells for coarse-graining and the correspondence between distribution functions and macroscopic variables. 
	\item Entropy:  $S(x) = k_B \text{log} |M(x)|,$ where $| \cdot |$ denotes the standard volume measure in phase space. Because of Fuzziness, in general, the (Boltzmann) entropy of a system is not exact. 
\end{itemize} 
We can translate WPH into the language of phase space: at one temporal boundary of space-time, the microstate of the universe $X_0$ lies inside  a particular macrostate $M_0$ that has low volume in phase space. 

Fuzziness is crucial for understanding the  vagueness and higher-order vagueness of  macrostates.   Without specifying the exact values (or exact ranges of values) of the C-parameters, the macrostates  have fuzzy boundaries: some microstates are borderline cases for certain macrostates, some are borderline borderline cases, and so on. The fuzzy boundary of $M_0$ illustrates  the existence of borderline microstates and higher-order vagueness.  There will be a precise identification of macrostates with sets of microstates only when we exactly specify the C-parameters (or their ranges). In other words, there is a precise partition of microstates on phase space into regions that are macroscopically similar (macrostates) only when we make some arbitrary choices about what the C-parameters are. In such situations, the WPH macrostate $M_0$ would correspond to an exact set $\Gamma_0$ on phase space, and the initial microstate has to be contained in $\Gamma_0$.  

However, proponents of the WPH do not  specify a precise set. A precise set $\Gamma_0$ would require more precision than is given in statistical mechanics---it requires the specific values of the coarse-grained cells and the specific correspondence with distribution functions. (In the standard quantum case discussed in Appendix, it also requires the precise cut-off threshold for when a superposition   belongs to a macrostate.)  The precise values of the C-parameters could be added to the theory to make WPH into a precise statement (which I call the \textit{Strong Past Hypothesis} in \cite{chen2018NV}). But they are nowhere to be found in the proposal, and rightly so.\footnote{For example, see descriptions of SWPH and WPH in \citealt{goldstein2001boltzmann}, \citealt{albert2000time}, and \citealt{carroll2010eternity}. }

Some choices of the C-parameters are clearly unacceptable. If the coarse-graining cells are too large, they cannot reflect the  variations in the values of macroscopic variables; if the coarse-graining cells are too small, they may not contain enough gas molecules to be statistically significant. Hence, they have to be macroscopically small but microscopically large (\cite{albert2000time} p.44(fn.5) and \cite{goldstein2019gibbs}).  However, if we were to make the parameters (or the ranges of parameters) more and more precise, beyond a certain point, any extra precision in the choice would seem completely arbitrary. They correspond to how large the cells are and which function is the correct one when defining the relation between temperature and sets of microstates. That does not seem to correspond to any objective facts in the world. (How large is large enough and how small is small enough?)  In this respect, the arbitrariness in precise C-parameters is quite unlike that in the fundamental dynamical constants. (In \cite{chen2018NV}, I discuss their differences in terms of a theoretical virtue called `traceability.') Moreover, not only do we lack precise parameters, we also lack a precise set of permissible parameters (hence no exact ranges of values for the C-parameters). There shouldn't be sharp boundaries anywhere. Suppose size $m$ is borderline large enough and size $n$ is determinately large enough. Small changes from $m$ will eventually get us to $n$, but it is implausible that there is a sharp transition from borderline large enough to determinately large enough. Similarly, there shouldn't be a sharp transition between borderline large enough to borderline borderline large enough, and so on.  That is higher-order vagueness.

Because of higher-order vagueness, \x{we need to take standard mathematical representation of WPH with a grain of salt}. The macroscopic variables---adjustable parameters in WPH---need to be coarse-grained enough to respect the vagueness. For example, we may \textit{represent} the temperature of $M_0$ as $10^{32}$ degrees Kelvin. But temperature does not have the exactness of real numbers. A more careful way to represent the vague temperature should be ``$10^{32}$-ish degrees Kelvin,'' where the ``-ish'' qualifier signifies that temperature is vague and the number $10^{32}$ is only an imperfect mathematical representation.\footnote{Thanks to Alan H\'ajek for discussions here.} Its  exactness is artificial. Hence, WPH should be characterized as a macrostate $M_0$ specified by  $S_0$-ish entropy, $T_0$-ish temperature, and so on. 

The vagueness here is appropriate, since  macroscopic variables only make sense when there are enough degrees of freedom (such as a large number of particles). In practice, however, such vagueness rarely matters:  there will be enough margins  such that to explain the thermodynamic phenomena, which are themselves vague, we do not need the extra exactness. The vagueness disappears \emph{for all practical purposes}. Nevertheless,  WPH is a genuine case of   \x{fundamental} nomic vagueness and it is a possibility to take seriously. 

\subsection{Summary}

I have suggested that fundamental laws of physics can be vague, and PH provides an example. It turns out that this may be an artifact of classical mechanics and can be naturally avoided in quantum mechanics \cite[sect. 4]{chen2018NV}. Hence, nomic exactness may be a metaphysically contingent feature of the universe that depends on the actual laws.

\section{Objectivity}

The final hallmark of laws I want to discuss is their objectivity. Laws are objective features of reality that do not depend on our beliefs or desires. They are mind-independent. However, the metaphysics of laws can make a difference to how we understand their objectivity. 

\subsection{Ratbag Idealism}

Humean Reductionism and non-Humeanism offer different understandings about the objectivity of laws. Let us focus on BSA and MinP. On BSA, assuming that we are relying on the right theoretical virtues and have appropriate access to the mosaic, the best summaries will be the true laws. There is a certain sense that, in principle, we \x{are guaranteed to be right.} On MinP, even if we rely on the correct theoretical virtues and the correct scientific methodology, we can still be mistaken about what the true laws are.  Epistemic guides are defeasible and fallible indicators for truth:  they do not guarantee that we find the true laws (although we may be rational to expect to find them). There are  fundamental, objective, and mind-independent facts about which laws govern the world, and we can be wrong about them. This is not a bug but a feature of MinP, symptomatic of the robust kind of realism that we endorse. For realists, this is exactly where they should end up; fallibility about the fundamental reality is a badge of honor. 

On BSA, since the best systematization is constitutive of lawhood, and what counts as best  may depend on us, then lawhood can become mind-dependent. In a passage about ``ratbag idealism,'' \cite{LewisHSD} discusses this worry and tries to offer a solution: 
\begin{quotation}
  The worst problem about the best-system analysis is that when we ask where the standards of simplicity and strength and balance come from, the answer may seem to be that they come from us. Now, some ratbag idealist might say that if we don't like the misfortunes that the laws of nature visit upon us, we can change the laws---in fact, we can make them always have been different---just by changing the way we think! (Talk about the power of positive thinking.) It would be very bad if my analysis endorsed such lunacy....
  
  The real answer lies elsewhere: if nature is kind to us, the problem needn't arise.... If nature is kind, the best system will be \textit{robustly} best---so far ahead of its rivals that it will come out first under any standards of simplicity and strength and balance. We have no guarantee that nature is kind in this way, but no evidence that it isn't. It's a reasonable hope. Perhaps we presuppose it in our thinking about law. I can admit that \textit{if} nature were unkind, and \textit{if} disagreeing rival systems were running neck-and-neck, then lawhood might be a psychological matter, and that would be very peculiar. (p.479)
\end{quotation}
For Lewis, the solution is conditionalized on the hope that nature is kind to us in this special way: the best summary of the world will be far better than its rivals. That may be a generous assumption, but it seems  consistent with scientific practice. 

In contrast, the worry about ratbag idealism does not arise on MinP; the objectivity of laws can be secured without appealing to the hope that nature is kind to us. On MinP, fundamental laws are what they are irrespective of our psychology and judgments of simplicity and informativeness. Even though the epistemic guides provide some guidance  for discovering and evaluating them, they do not guarantee arrival at the true fundamental laws. Moreover, changing our psychology or judgments will not change which facts are fundamental laws. Hence, MinP respects our conviction about the objectivity and mind-independence of fundamental laws.

\subsection{A Package Deal}

Despite the difference regarding the metaphysical objectivity of laws, there is a convergence in methodological principles underlying MinP and BSA, which may be traced back to epistemic principles that are somewhat mind-dependent.  In \S7.3, we discuss the epistemic version of ratbag idealism. 

We have mainly focused on fundamental laws. To be sure, they are related to the fundamental material ontology (fundamental entities and their properties). On MinP, we regard both fundamental laws and fundamental material ontology as metaphysical primitives and evaluate them in a package. In this respect,  MinP is similar to  Loewer's Package Deal Account (PDA),  a descendent of BSA that regards both as co-equal elements of a package deal \citep{loewer2020package, loewer2021fire}, but they also have significant differences. 

On  PDA, we look for the best systematization in terms of a package of laws and (material) ontology; the package is  supervenient on the actual world.  Thus, fundamental laws and fundamental ontology enter the discussion in the same way, at the same place, and on the same level. MinP shares this feature, although fundamental ontology and fundamental laws are merely discovered by us and not made by us or dependent on us. On PDA, given the actual world (of which we have very limited knowledge), we evaluate different packages of laws + ontology, and we evaluate them based on our actual scientific practice. Hence, there will be some degree of relativism. Relative to different scientific practice or a different set of scientists, the judgement as to the actual laws + ontology would have been different. Consequently on PDA, fundamental laws and fundamental ontology are dependent on us in a significant way.  

On MinP, we may use the best package-deal  systematization as a guide to discover the laws and ontology; given the actual world (of which we have very limited knowledge), we evaluate different packages of  laws + ontology, and we evaluate them based on our actual scientific practice. Hence, there will be some degree of uncertainty. Relative to different scientific practice or a different set of scientists, the judgement as to the actual laws + ontology would have been different. Still, what they are is metaphysically independent of our belief and practice. 

Here I quote a passage from \cite{loewer2021fire}. Although we disagree with him on the metaphysics, we agree on how the enterprise of physics should be understood:
\begin{quotation}
    The best way of understanding the enterprise of physics is that it begins, as Quine says, ``in the middle'' with the investigation of the motions of macroscopic material objects e.g., planets, projectiles, pendula, pointers, and so on. Physics advances by proposing theories that include laws that explain the motions of macroscopic objects and their parts. These theories may (and often do) introduce ontology, properties/relations, and laws beyond macroscopic ones with which it began and go onto to posit laws that explain their behaviors...... The ultimate goal of this process is the discovery of a theory of everything (TOE) that specifies a fundamental ontology and fundamental laws that that cover not only the motions of macroscopic objects with which physics began but also whatever additional ontology and quantities that have been introduced along the way. (pp. 30-31)
\end{quotation}
From the perspective of MinP, this is an excellent description of how fundamental laws and ontology are discovered---in a package.

\subsection{Humeanism and non-Humeanism}

For non-Humeans who endorse methodological principles as epistemic guides, the problem of ratbag idealism may show up in the epistemology of laws. \cite[\S5.6]{hall2009humean} suggests that, facing the problem that the simplicity criterion in the BSA is too subjective, Lewis and other Humeans can ``perform a nifty judo move.'' If non-Humeans regard simplicity as an epistemic guide to laws, ``central facts of normative epistemology are \textit{also} up to us.'' Hall argues that this is more objectionable than the ratbag idealism of BSA. A defender of BSA may reasonably embrace ratbag idealism and take laws to be pragmatic tools to structure our investigation of the world. With that viewpoint, we can expect that what laws are is somewhat up to us.  However, there is no reason on non-Humeanism why epistemological and normative facts should be up to us. So the non-Humeans might face a worse problem of ratbag idealism. 

My analysis in \S5 suggests that both Humeans and non-Humeans need to adopt fairly strong epistemological principles such as PNS. On Humeanism, there are two independent appeals to simplicity (among other theoretical virtues). The metaphysical analysis of laws with BSA requires laws be no more complicated than the mosaic. But not all mosaics support simple laws; in fact, many metaphysically possible mosaics may not have any regularity that deserves the label of laws. To believe, on BSA, that  simple laws exist is to believe that the actual  mosaic is very special.  The epistemic guide to simple laws serves as an epistemic and normative restriction of possible candidate mosaics Humeans ought to consider. Why the Humean mosaic should be so nice on BSA is basically the same question as why fundamental constraints should be so simple on MinP. So on both BSA and MinP, epistemological and normative facts are up to us. 

This reveals a deeper difference between Humeanism and non-Humeanism. On MinP, we should assume that certain fundamental facts of the world are simple. On BSA, in contrast, we should assume that certain superficial facts of the world (best-system laws), grounded in a complex fundamental reality, are simple. The assumption, it seems to me, is much more believable on MinP than on BSA. It is much easier to believe that nature at some deep level is very simple. It is much more difficult to believe that nature at some deep level is very complicated in a certain way to give rise to a simple appearance. Of course, this would not persuade the committed Humeans, for presumably they are willing to accept the consequence. However, for many people on the fence or coming to this debate for the first time, the choice between Humeanism and non-Humeanism should be clear.\footnote{I am indebted to discussions with Tyler Hildebrand and Boris Kment about this point.}


\subsection{Summary}

Metaphysical objectivity is a feature of  laws on non-Humeanism such as MinP but not on Humean views such as BSA. However, on both theories the epistemological principles may be somewhat up to us. 


\section{Conclusion}

Laws occupy a central place in a systematic philosophy of the physical world. They can be regarded as fundamental facts that govern the universe by constraining its physical possibilities. With this minimal primitivist account, one accepts that laws transcend the concrete physical reality they govern, but need not presume a fundamental direction of time or require a fundamental ontology of universals, dispositions, or counterfactuals. The account allows one to contemplate a variety of candidate fundamental laws, and to understand the marks of the nomic as arising from methodological and epistemological principles. It is not the only viewpoint available, but it is one I recommend to anyone looking for an account that illuminates metaphysics but is not unduly constrained by it.

\section*{Acknowledgements}

I am indebted to Sheldon Goldstein for invaluable discussions about laws of physics over the years. This volume is connected to several issues explored in our joint paper on MinP. I thank Shelly Yiran Shi and Bosco Garcia for their research assistance and helpful comments on several earlier drafts. For discussions and encouragements, I am also grateful to Emily Adlam, David Albert, Jeffrey Barrett, Craig Callender, Eugene Chua, Christopher Dorst,  Nina Emery, Ned Hall, James Hartle, Tyler Hildebrand,  Boris Kment, Marc Lange, Barry Loewer, Tim Maudlin, Chris Meacham, Kerry McKenzie, John Roberts, Carlo Rovelli, Charles Sebens, Elliott Sober, Eric Watkins, and James Owen Weatherall. This project is supported by an Academic Senate Grant from the University of California, San Diego.


\bibliography{test}


\end{document}